\documentclass[english,nofootinbib]{revtex4}
\usepackage[T1]{fontenc}
\usepackage[latin9]{inputenc}
\usepackage{amsmath}
\usepackage{graphicx}
\usepackage{amssymb}

\makeatletter

\providecommand{\tabularnewline}{\\}

\@ifundefined{textcolor}{}
{%
 \definecolor{BLACK}{gray}{0}
 \definecolor{WHITE}{gray}{1}
 \definecolor{RED}{rgb}{1,0,0}
 \definecolor{GREEN}{rgb}{0,1,0}
 \definecolor{BLUE}{rgb}{0,0,1}
 \definecolor{CYAN}{cmyk}{1,0,0,0}
 \definecolor{MAGENTA}{cmyk}{0,1,0,0}
 \definecolor{YELLOW}{cmyk}{0,0,1,0}
 }

\makeatother

\usepackage{babel}

\begin{document}

\title{Infrared singularities in Landau gauge Yang-Mills theory}

\author{Reinhard Alkofer, Markus Q. Huber and Kai Schwenzer}

\address{Institut für Physik, Karl-Franzens Universität Graz, Universitätsplatz
5, 8010 Graz, Austria}
\begin{abstract}
We present a more detailed picture of the infrared regime of Landau
gauge Yang-Mills theory. This is done within a novel framework that
allows one to take into account the influence of finite scales within
an infrared power counting analysis. We find that there are two qualitatively
different infrared fixed points of the full system of Dyson-Schwinger
equations. The first extends the known scaling solution, where the
ghost dynamics is dominant and gluon propagation is strongly suppressed.
It features in addition to the strong divergences of gluonic vertex
functions in the previously considered uniform scaling limit, when
all external momenta tend to zero, also weaker kinematic divergences,
when only some of the external momenta vanish. The second solution
represents the recently proposed decoupling scenario where the gluons
become massive and the ghosts remain bare. In this case we find that
none of the vertex functions is enhanced, so that the infrared dynamics
is entirely suppressed. Our analysis also provides a strict argument
why the Landau gauge gluon dressing function cannot be infrared divergent.
\end{abstract}
\maketitle

\section{Introduction}

The combined effort of functional approaches and lattice gauge theory
led in the past years to a comprehensive picture of the qualitative
features of the infrared (IR) limit of Yang-Mills theory in Landau
gauge \cite{Alkofer:2000wg,Fischer:2006ub,Bowman:2007du}. This qualitative
information is encoded in a set of IR power laws for the Green functions
of the theory. These can incorporate important aspects of the confinement
mechanism for gluons within the scenarios of Kugo-Ojima \cite{Kugo:1979gm}
and Gribov-Zwanziger \cite{Gribov:1977wm} and serve as a basis for
the inclusion of matter fields and the challenging problems of quark
confinement \cite{Alkofer:2008tt,Schwenzer:2008vt,Braun:2007bx}
and spontaneous chiral symmetry breaking \cite{Fischer:2006ub,Alkofer:2006gz}.
\\
Previous studies of the Dyson-Schwinger equations (DSEs) \cite{Roberts:1994dr,Alkofer:2000wg,Fischer:2006ub}
suggested that the qualitative aspects of the Yang-Mills IR fixed
point structure in Landau gauge are already known \cite{vonSmekal:1997is,Zwanziger:2001kw,Lerche:2002ep,Alkofer:2004it}.\textbf{
}It had been shown that when the DSEs and functional renormalization
group equations are combined, within the class of \emph{scaling} fixed
points, where no masses are dynamically generated, there exists a
unique solution \cite{Fischer:2006vf}. This conclusion was possible
by the complementary constraints obtained from the two different hierarchies
of equations and we will show here that, taking into account the constraints
provided by the existence of the skeleton expansion, this is already
implied by the DSE system alone. Yet, analog to the propagators,
which depend on a single external scale, it was expected that the
IR limit of a Green function is uniquely determined by a single IR
scaling law. In the case of the vertices, however, the situation is
more diverse. Besides the appearance of a multitude of different tensor
structures, the corresponding form factors are also functions of several
distinct momenta. Therefore, there are in general different combinations
of momenta that can become soft and lead to IR divergences. In present
studies the implicit assumption was that vertices become IR divergent
if and only if all scales go to zero. This led to general results
for the IR scaling of arbitrary vertices in this uniform limit \cite{Alkofer:2004it}
that were later extended to arbitrary dimension \cite{Huber:2007kc}.
Here we show that this picture - although qualitatively correct -
needs to be refined in the sense that there are additional kinematic
singularities that characterize the IR-regime. \\
It is known since the early work of Taylor \cite{Taylor:1971ff}
that the IR limit of the ghost-gluon vertex in Landau gauge is not
influenced by radiative corrections when an external ghost momentum
vanishes. This result is used as the starting point for various IR
analyses and is explicitly confirmed by our analysis. In contrast
we find a kinematic singularity in the 3-gluon vertex. Although this
kinematic singularity which reflects the sensitivity to ultrasoft
gluon exchange is rather mild, $\sim\left(p^{2}\right)^{1-2\kappa}$
with $\kappa\gtrsim0.5$ \cite{Zwanziger:2001kw,Lerche:2002ep},
it could be conceptually important. First, due to to the parametrically
larger support in loop integrals kinematic singularities should have
a sizable impact on the quantitative results for Yang-Mills Green
functions. Further, it has been argued recently that the Slavnov-Taylor
identity for the 3-gluon vertex suggests that the gluon propagator
is IR divergent \cite{Boucaud:2007hy}. This argument relied on the
assumption that the 3-gluon vertex is finite when only a single momentum
vanishes. Our results show that this is not the case for the scaling
solution and thereby the corresponding conclusion cannot be drawn.
In addition we show here that the DSEs for the gluonic vertices lead
without any approximation or assumption to the condition that the
gluon dressing function \emph{cannot be} IR divergent \cite{Fischer:2006vf}
as argued by Mandelstam \cite{Mandelstam:1979xd}. Finally and even
more important, the techniques developed here allow one to study the
quark sector of QCD, where a corresponding kinematic singularity in
the quark-gluon vertex is induced \cite{Alkofer:2008tt}. In contrast
to the gauge vertex where the kinematic singularity is induced semi-perturbatively,
cf. \cite{Alkofer:2008dt}, the one in the quark-gluon vertex is
much stronger divergent due to a self-consistent enhancement mechanism.
Thereby it induces long ranged gluonic interactions that can confine
quarks \cite{Alkofer:2008tt,Schwenzer:2008vt}. \\
In contrast to the assumptions of present scaling analyses the
DSE system entails also the possibility of dynamical mass generation.
In addition to the IR \emph{scaling} solution a \emph{decoupling}
solution where the gluons become massive has recently been suggested
\cite{Boucaud:2008ji,Aguilar:2008xm,Fischer:2008uz}, see also \cite{Cornwall:1981zr,Dudal:2008sp}.
This solution is also observed in recent lattice simulations \cite{Bowman:2007du,Cucchieri:2007rg}.
Whereas there is only a single scaling solution the decoupling case
seems to allow a whole family of solutions characterized by the IR
gluon mass that continuously connects with the scaling solution. These
different solutions are obtained depending on the chosen boundary
conditions for the functional equations \cite{Fischer:2008uz}, cf.
also \cite{Boucaud:2008ji}. In case of the decoupling solution certain
IR Green functions are not dominated by IR modes but by finite scales
of the order of the characteristic scale $\Lambda_{QCD}$. We extend
the power counting formalism in order to take this possibility into
account in our IR analysis. Thereby we confirm that the decoupling
is compatible with the vertex equations and indeed presents a second
solution of the full DSE system. However, we find that within the
decoupling solution none of the vertices is IR enhanced. Correspondingly,
no long ranged interaction is reflected in the Green functions of
the fundamental local degrees of freedom.\\
This article is closely related to another article \cite{Alkofer:2008dt}
on the IR limit of Yang-Mills theory and the behavior of the vertices.
There we provide explicit, analytic solutions for the 3-point vertices
in a semi-perturbative approximation and confirm and elaborate on
the results discussed here.

\section{Generalized infrared fixed point analysis\label{sec:IR-analysis}}

In this section we present the general formalism that allows one to
study the non-perturbative IR scaling behavior of Green functions.
In particular, we take into account that the IR fixed points can be
influenced by large scales, like external momenta of Green functions
that remain finite in the IR limit and can lead to kinematic singularities,
or dynamically generated masses. Our analysis is based on the system
of Dyson-Schwinger equations that describe the complete non-perturbative
dynamics of the considered theory. The equations for arbitrary vertex
functions can be derived algorithmically as shown in \cite{Alkofer:2008nt}
and as has been implemented in the Mathematica package \emph{DoDSE}.
The algorithm can be represented by diagrammatic replacement rules
applied to the underlying equations for the 1-point functions and
is sketched in Appendix \ref{sec:DSE-derivation}.

\begin{figure}
\includegraphics[scale=0.65]{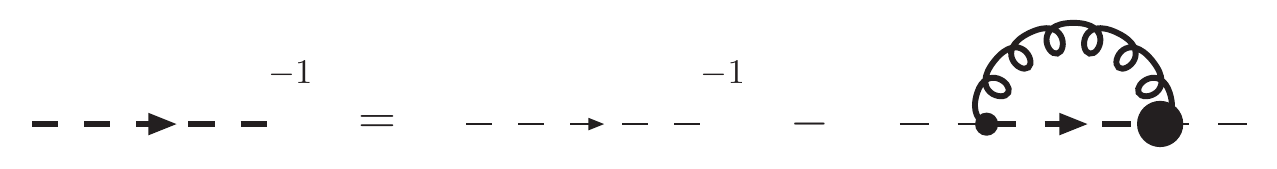}

\caption{\label{fig:ghost-DSE}DSE for the ghost-propagator. Thin and thick
lines and dots represent bare and proper propagators and vertices.
All proper Green functions are amputated and the external lines are
shown only for illustration purposes.}

\end{figure}

\begin{figure}
\includegraphics[scale=0.65]{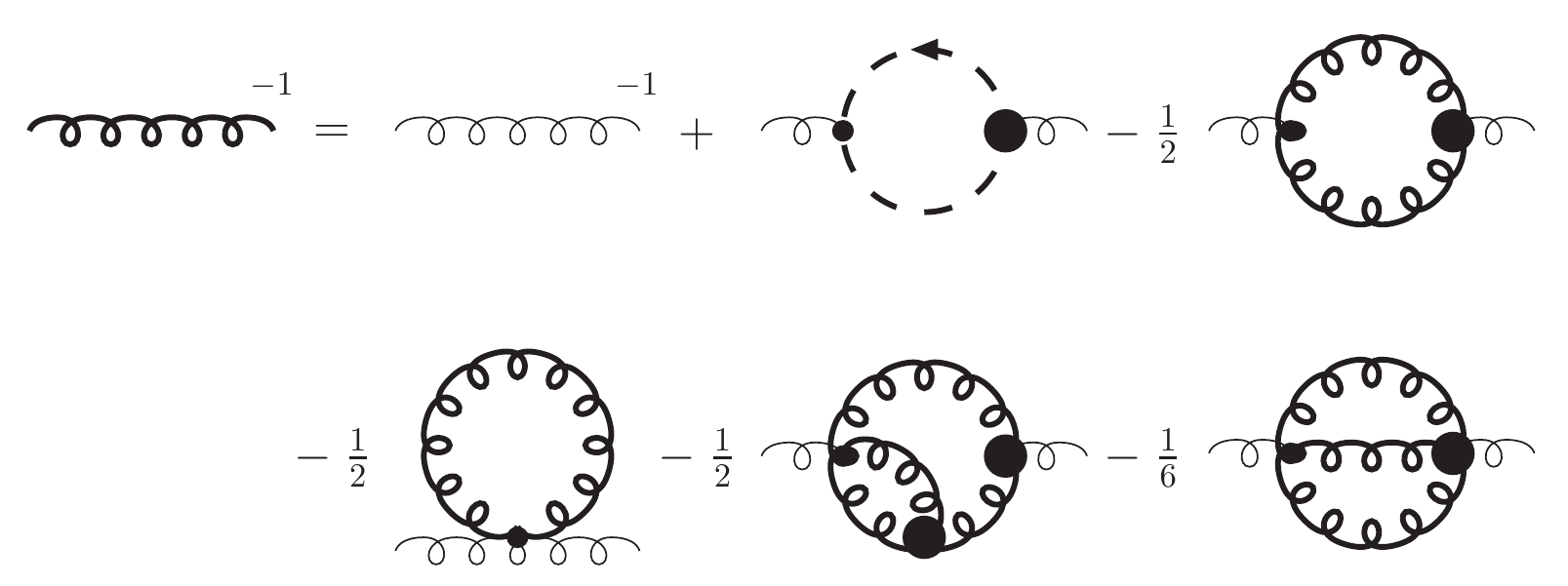}

\caption{\label{fig:gluon-DSE}DSE for the gluon-propagator.}

\end{figure}

\begin{figure}
\includegraphics[scale=0.65]{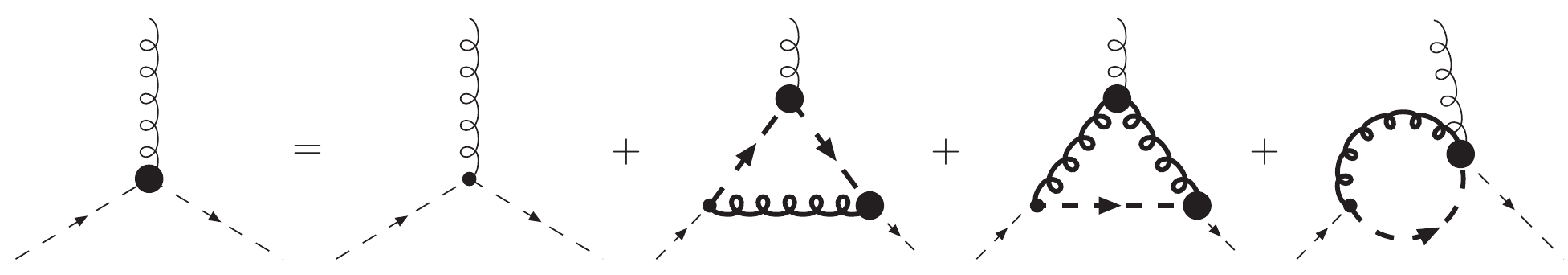}

\caption{\label{fig:ghost-gluon-DSE-1}First version of the DSE for the ghost-gluon
vertex. The first order in a skeleton expansion consists of the first
three diagrams on the right-hand side. The fourth diagram, which features
the 1PI ghost-gluon scattering kernel, contributes only at higher
order.}

\end{figure}

\begin{figure}
\includegraphics[scale=0.85]{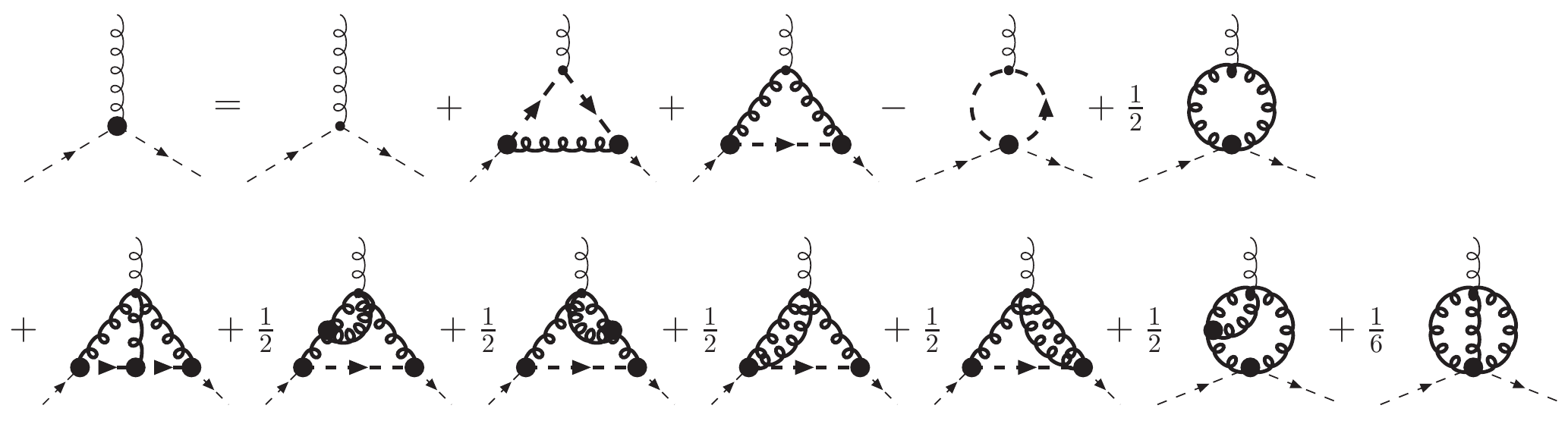}

\caption{\label{fig:ghost-gluon-DSE-2}Second version of the DSE for the ghost-gluon
vertex. Here the leading order in the skeleton expansion of the right
hand side consists of the first three diagrams in the first and second
lines, respectively.}

\end{figure}

\begin{figure}
\includegraphics[scale=0.85]{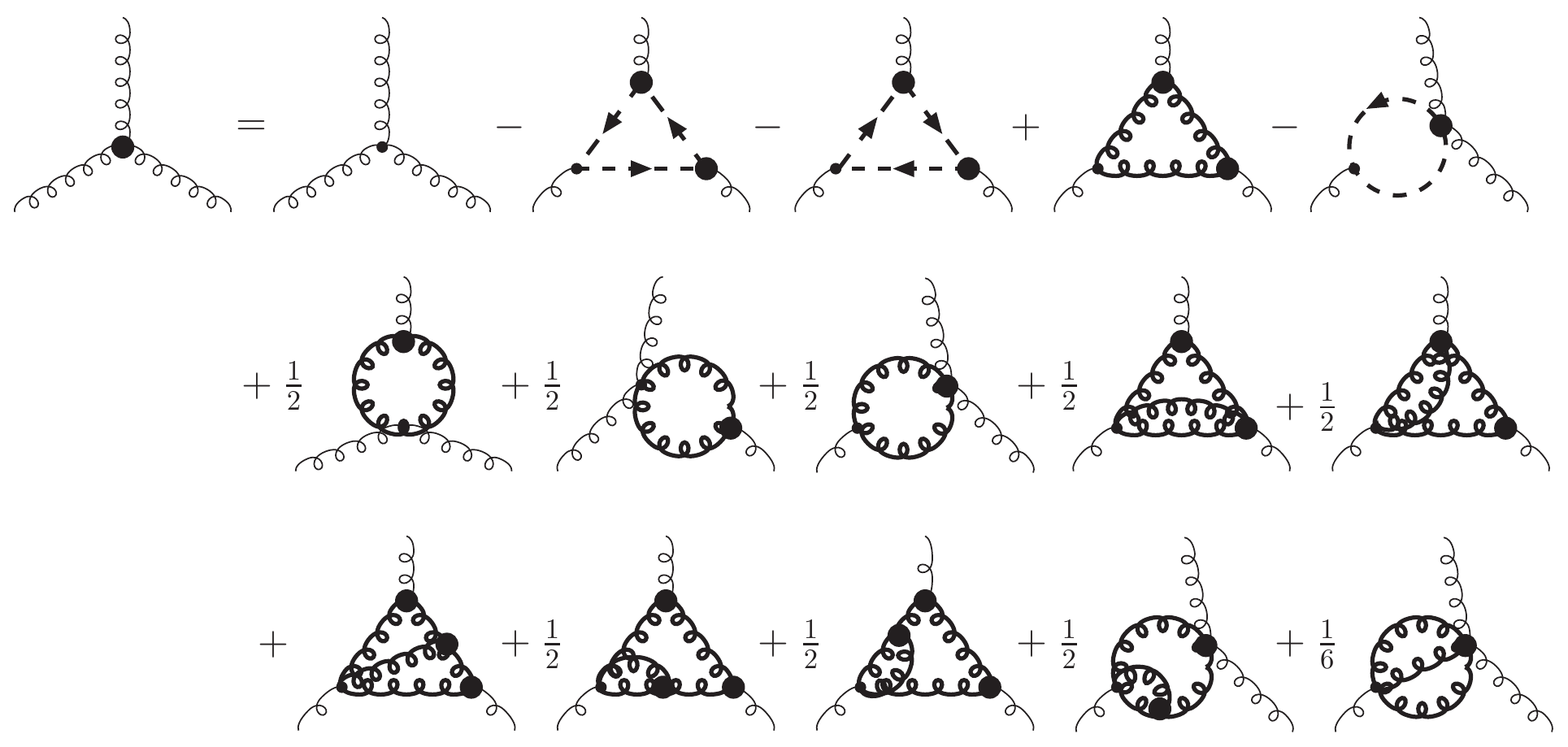}

\caption{\label{fig:3-gluon-DSE}Full DSE for the 3-gluon vertex. To leading
order in a skeleton expansion the last graphs in the first respectively
third line are absent. The restriction to the 1-loop graphs reduces
to the leading order in $\alpha_{s}$ when dressed Green functions
are replaced by their tree-level expressions.}

\end{figure}

\begin{figure}
\includegraphics[scale=0.8]{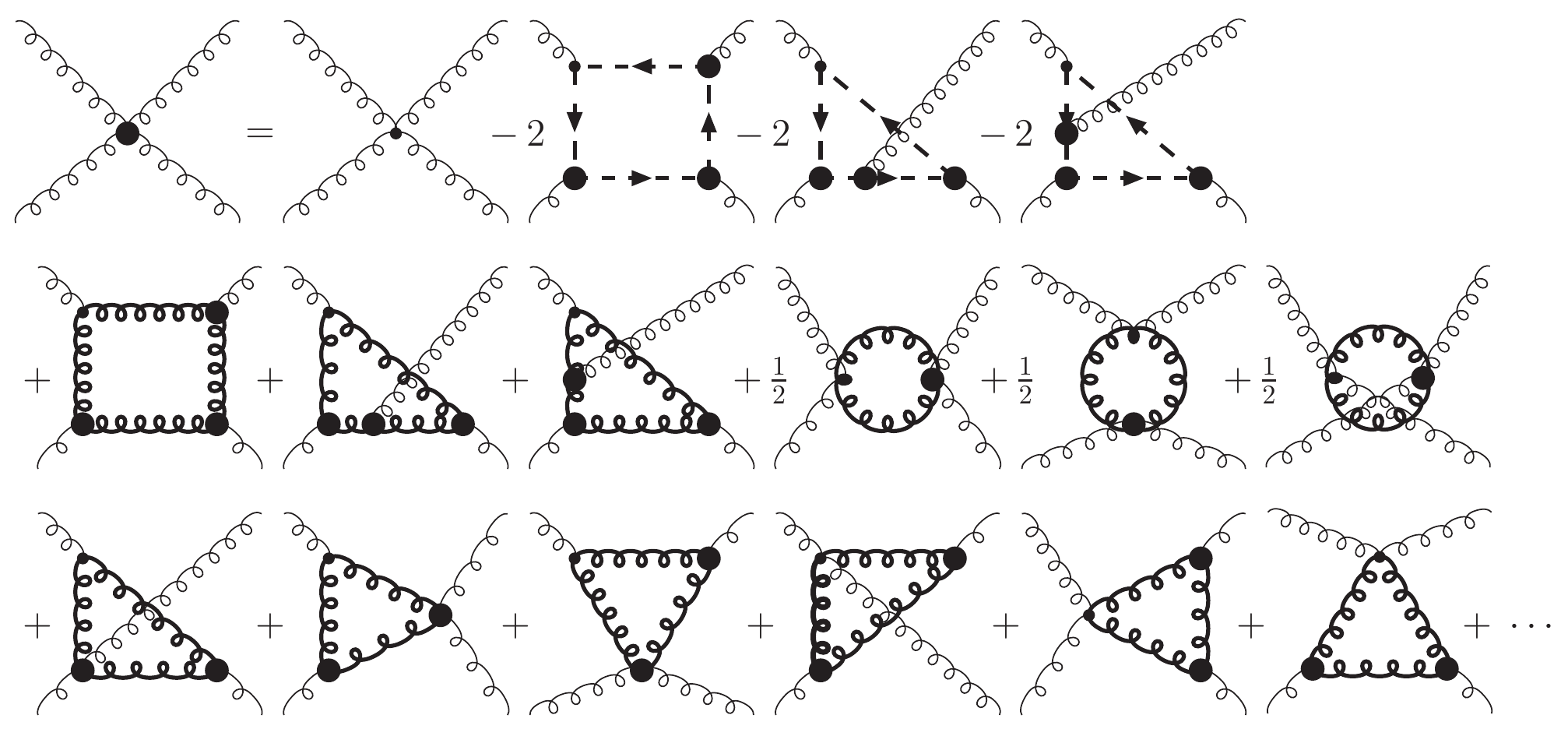}

\caption{\label{fig:4-gluon-DSE}DSE for the 4-gluon vertex - for conciseness
we only show the 1-loop contribution in the skeleton expansion. This
reduces to the leading order in $\alpha_{s}$ when dressed Green functions
are replaced by their tree-level expressions.}

\end{figure}

\begin{figure}
\begin{minipage}[t]{0.35\columnwidth}%
\includegraphics[scale=0.75]{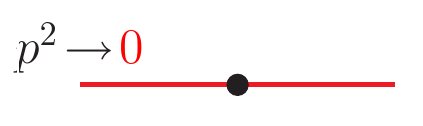}%
\end{minipage}%
\begin{minipage}[t]{0.65\columnwidth}%
\includegraphics[scale=0.75]{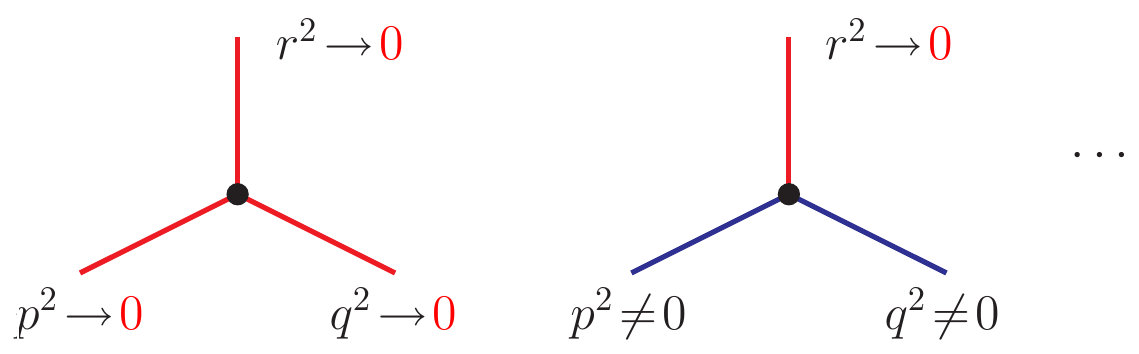}%
\end{minipage}

\caption{The unique IR singular kinematic configuration of the propagator (left)
and the different IR singular kinematic configurations of the 3-point
functions (right).\label{fig:3-point-kinematics}}

\end{figure}

\begin{figure}
\includegraphics[scale=0.75]{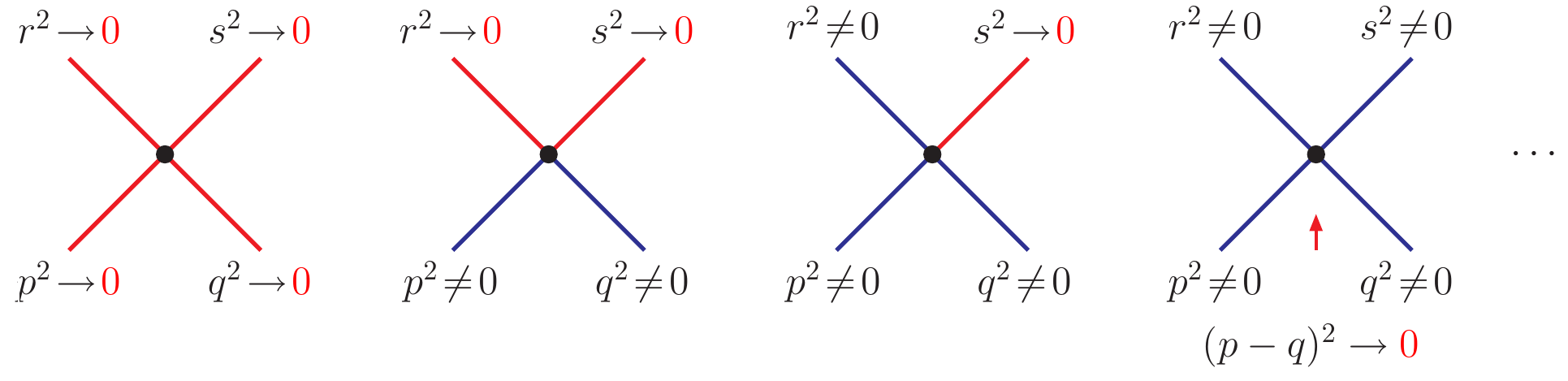}

\caption{IR singular kinematic configurations of the 4-point functions. In
this case there can even be kinematic singularities when all external
momenta are hard.\label{fig:4-point-kinematics}}

\end{figure}

The equations for the propagators and primitively divergent vertices
of Landau gauge Yang-Mills theory for instance are given in figs.
\ref{fig:ghost-DSE} to \ref{fig:4-gluon-DSE}. As can be seen, the
DSEs for the primitively divergent Green functions are in general
not closed. Instead they are part of an infinite hierarchy of coupled
equations and thereby it might seem hopeless to make any definite
statements about existence and uniqueness of possible fixed point
solutions. However, as has been demonstrated in \cite{Alkofer:2004it},
this is even possible on the level of a mere IR power counting analysis.
The \emph{working hypothesis} of such an analysis is that in the IR
regime the theory can still be described by the field content of the
classical action. It involves in the present case as a tool the concept
of a skeleton expansion of higher order correlation functions in terms
of skeleton diagrams. This means that all vertex functions that do
not appear in the Lagrangian are replaced by a loop expansion that
features only primitively divergent vertices and propagators. Consequently
the system of equations is closed, as it consists only of the DSEs
of the primitively divergent vertex functions. The difference compared
to a perturbative expansion of these higher vertex functions is, however,
that all the primitively divergent Green functions in the expansion
loops are dressed. Since the DSEs employed here are based on one-particle
irreducible (1PI) vertices, graphs that involve vertices that are
not primitively divergent do not contribute to leading order in the
skeleton expansion. E. g. for the vertices of Landau gauge Yang-Mills
theory the leading order in the skeleton expansion is discussed in
the captions of figs. \ref{fig:ghost-gluon-DSE-1} to \ref{fig:4-gluon-DSE}.
The skeleton expansion yields a closed DSE system for the primitively
divergent Green functions but involves graphs of arbitrary loop order.
In the framework of a power counting analysis it is possible to assess
the IR scaling of all these graphs since higher orders can be generated
by simple extension rules \cite{Alkofer:2004it}. As discussed in
more detail below, such extensions by additional loops should in particular
not increase the degree of IR divergences since higher order diagrams
would otherwise be more and more divergent and the description in
terms of the underlying degrees of freedom would break down. Correspondingly
the lowest order in the expansion already entails in general the leading
IR scaling. The solutions obtained with a skeleton expansion are by
construction solutions of the full system of DSEs, where the scaling
of higher order Green functions is already determined by the leading
order skeleton graphs. Note that the skeleton expansion does not miss
additional solutions where the non-trivial behavior is triggered by
$n$-point functions, $n>4$ \cite{Schwenzer:2008vt}. In the case
of scaling fixed points the IR analysis has already been performed
without the tool of the skeleton expansion in \cite{Huber:2009wh,Fischer:2006vf}.
Furthermore, the equivalence between the constraints obtained from
the skeleton expansion and constraints derived from functional renormalization
group equations has been shown in \cite{Huber:2009wh}. Hence the
use of the skeleton expansion is well justified.\\
Whether a solution, where the underlying degrees of freedom are
still valid in the IR regime, actually exists can only be decided
by an explicit analysis of the non-perturbative dynamics. Once found,
a corresponding solution exists then independently of the skeleton
expansion and directly verifies the working hypothesis. Yet, our analysis
does not exclude other possible solutions where this is not the case.
A consequence of our working hypothesis is that there should be no
singularities in the external momenta in two- and three-point functions
as long as all scales are finite. Such a singularity arises e.g. in
the perturbative running of the strong coupling in the form of the
Landau pole, but in this case it merely reflects the insufficient
dynamical treatment. Yet, in a fully non-perturbative analysis it
would signal the breakdown of the description in terms of the fundamental
gauge degrees of freedom. In contrast, it is for instance generally
expected that the physical degrees of freedom in Yang-Mills theory
are glueball bound states which are reflected as singularities in
higher gluonic $n$-point functions with $n\geq4$. These singularities
arise at finite invariant momenta corresponding to the masses of the
glueball states. In Euclidean space these poles are on the negative
half axis of the squared momentum variable and do not interfere with
our analysis. \\
The motivation for an IR analysis is that certain theories, including
in particular the important case of gauge theories, are scale invariant
at the classical level. Yet, the quantum theory can dynamically generate
a scale $\Lambda$ via dimensional transmutation. Whereas e.g. in
a perturbative analysis of Yang-Mills theory the corresponding scale
$\Lambda_{QCD}$ is renormalization scheme dependent but otherwise
uniquely defined by the Landau pole, non-perturbatively there is no
unique definition and it can be chosen as an arbitrary hadronic scale.
For the scaling solution of Landau gauge Yang-Mills theory \cite{vonSmekal:1997is},
discussed in more detail below, it could for instance be identified
with the scale where the gluon propagator reaches its maximum. The
leading behavior of all Green functions \emph{far below} the scale
$\Lambda$ should by renormalization group arguments be described
by a power law scaling with appropriate IR exponents. Although such
a power law form is obtained analytically in conformal field theories
and has been observed in scale free regimes of simpler models, for
Yang-Mills theory the existence of an IR power law scaling has to
our knowledge not been rigorously proven, yet, and \emph{presents
the main assumption we make in our analysis}. To be precise, in this
work we determine the IR solutions of Yang-Mills theory that feature
IR power law scaling but cannot exclude other possible solutions that
involve an IR momentum dependence with a more complicated analytic
structure. In particular it is conceivable that in addition to a power
law scaling there are additional logarithmic dependencies, like it
is the case in the UV regime of the theory. We do not explicitly take
into account this possibility, but we note that such additional logarithmic
dependencies would only refine the scaling and not change the dominant
power law structure. Thereby the qualitative results of this analysis
remain unaltered. \\
Whereas the propagators depend only on a single external momentum
scale and correspondingly have a unique IR behavior, the vertices
involve several momenta and can feature different IR power laws in
different kinematic sections. To expose the IR behavior, we perform
a tensor decomposition of a given proper Green function $\Gamma_{v}$
with appropriate Lorentz indices $\mu_{i}$ and internal indices
$a_{j}$ in some tensor basis $T_{t}$ that is analytic in the momenta

\begin{align}
\left(\Gamma_{v}\right)_{\mu_{1}\cdots\mu_{m}}^{a_{1}\cdots a_{n}}\left(q_{1},\cdots,q_{n}\right) & =\sum_{t}\Gamma_{v,t}\left(q_{1}^{2},q_{1}\!\cdot\! q_{2},\cdots,q_{n}^{2}\right)\left(T_{t}\right)_{\mu_{1}\cdots\mu_{m}}^{a_{1}\cdots a_{n}}\left(q_{1},\cdots,q_{n}\right)\label{eq:tensor-dec}\end{align}
so that the corresponding dressing functions $\Gamma_{v,t}$ depend
only on scalar arguments. Under the assumption of IR scaling, Green
functions can in principle have distinct IR power laws whenever any
combination of the external momenta vanishes. The dressing functions
generally take this power law scaling form only in the deep IR regime
$p_{i}^{2}\ll\Lambda$. For intermediate momenta of the order of $\Lambda$
they can involve a more intricate structure and in the UV region they
feature a different asymptotic scaling behavior, that is e.g. in Yang-Mills
theory given by the perturbative logarithmic RG running. Nevertheless,
the full dressing functions can be formally decomposed into a sum
of terms that include the individual IR power laws in the different
kinematic regimes \begin{align}
\Gamma_{v,t}\left(q_{1}^{2},q_{1}\!\cdot\! q_{2},\cdots,q_{n}^{2}\right) & =\sum_{l}\gamma_{v,t}^{l}\left(q_{1}^{2},q_{1}\!\cdot\! q_{2},\cdots,q_{n}^{2}\right)\left(\frac{p_{l}^{2}\left(q_{1}^{2},\cdots,q_{n}^{2}\right)}{\Lambda^{2}}\right)^{\delta_{v,t}^{l}}\,.\label{eq:general-scaling}\end{align}
In the following we explain this expression in detail. Here, the
set of functions $p_{l}^{2}$ defines the scaling variables in the
different kinematic limits $l$ which are chosen analytic in all arguments
and identical for all Green functions. The $\delta_{v,t}^{l}$ are
the corresponding power law exponents. In the IR limit that the scaling
variable vanishes, the corresponding function $\gamma_{v,t}^{l}$
in the decomposition eq. (\ref{eq:general-scaling}) is defined such
that it depends only on finite momentum ratios\begin{equation}
\gamma_{v,t}^{l}\left(q_{1}^{2},q_{1}\!\cdot\! q_{2},\cdots,q_{n}^{2}\right)\xrightarrow[p_{l}^{2}\rightarrow0]{}\gamma_{v,t}^{l}\left(\left\{ \frac{q_{l_{j}}^{2}}{p_{l}^{2}}\right\} \right)\:.\end{equation}
where the scaling variable $p_{l}$ depends only on the momenta $q_{l_{j}}$
that vanish in the kinematic limit. In particular if the scaling variable
depends only on a single external momentum, $\gamma_{v,t}^{l}$ becomes
constant in the IR limit. In general, the IR exponents $\delta_{v,t}^{l}$
also depend on the specific tensor. In our analysis we will not distinguish
between the different tensor structures and are only interested in
the IR exponents of the \emph{most singular} dressing functions $\delta_{v}^{l}\equiv\min_{t}\left(\delta_{v,t}^{v}\right)$
which will dominate in the IR.\\
We should stress, that the additive decomposition eq. (\ref{eq:general-scaling})
which explicitly separates the different IR divergences of a dressing
function into different summands does not represent an IR approximation
and is exact on all momentum scales. Yet, it is surely far from unique.
This ambiguity is reflected in the functions $\gamma_{v,t}^{l}$ which
include the proper mid and high momentum behavior and also depend
on the choice of the arbitrary scale $\Lambda$. This means that the
explicit IR power law is compensated in the functions $\gamma_{v,t}^{l}$
outside of the IR region. As discussed above, for $n$-point functions
with $n\geq4$ they may involve singularities when all combinations
of external scales are finite but they are by definition regular whenever
external scales vanish. The only requirement on the partition is the
complete decomposition of possible IR singularities. For the determination
of the IR power law exponents, however, the detailed form of the decomposition
at finite scales is irrelevant as discussed below. \\
An important IR limit is determined by a \emph{uniform} scaling
variable that is given by a function $p_{u}$ that fulfills

\begin{equation}
p_{u}^{2}\left(q_{1}^{2},\cdots,q_{n}^{2}\right)\to0\qquad\Leftrightarrow\qquad q_{1},\cdots,q_{n}\to0\quad\wedge\quad q_{1}^{2}/p_{u}^{2},\cdots,q_{n}^{2}/p_{u}^{2}\quad\mathrm{constant}\:.\label{eq:uniform-variable}\end{equation}
Such a function is for instance provided by the {}``Euclidean norm''

\begin{equation}
p_{u}^{2}\left(q_{i}^{2}\right)\equiv\sum_{i}q_{i}^{2}\:.\end{equation}
Since in this scaling limit all external momenta scale uniformly to
zero this presents the fully scale-invariant case. Previous studies
assumed that all Green functions are finite in other kinematic limits
and thereby have only considered this {}``conformal''%
\footnote{The term conformal refers in our context only to the scale invariance
property but not the additional symmetries included in the conformal
group. %
} case \cite{Alkofer:2004it,Fischer:2006vf,Huber:2007kc}. We will
discuss this important approximation in section \ref{sec:Conformal-scaling}.\\
In general, however, there are various different kinematic limits
where the $p_{l}$ can also depend on any subset of the $q_{i}$.
As we will argue below in section \ref{sec:Soft-singularities}, in
Yang-Mills theory such a more general IR behavior is realized that
involves kinematic singularities according to eq. (\ref{eq:general-scaling})
when only a subset of the external momenta vanishes. We will denote
those external momenta that scale to zero as \emph{soft} and those
that stay fixed when the IR limit is taken as \emph{hard} momenta.
In particular, all soft external momenta are assumed to be much smaller
than $\Lambda$ in order to ensure the applicability of IR power laws.
By momentum conservation it is impossible that exactly $n-1$ momenta
of an $n$-point function become soft, but e.g. all other cases when
$1\leq i\leq n-2$ momenta tend to zero can define a separate scaling
limit. The simplest such kinematic limit is when a single external
momentum becomes soft and all others remain hard where the corresponding
scaling variable in eq. (\ref{eq:general-scaling}) is $p_{i}^{2}\equiv q_{i}^{2}$.
For each distinct external field this presents a different limit described
by a separate power law exponent. These present the only kinematic
limits in case of 3-point functions, as illustrated in fig. \ref{fig:3-point-kinematics}.
As shown in fig. \ref{fig:4-point-kinematics}, for the 4-point functions
there are already two distinct limits when one or two external legs
are soft. In addition, there can even be IR singularities for 4-point
functions when all external momenta are hard but differences of momenta
become small corresponding to the exchange of a soft momentum in the
respective intermediate channel. For higher order Green functions
the number of IR exponents rises further.


\begin{figure}
\includegraphics[scale=0.75]{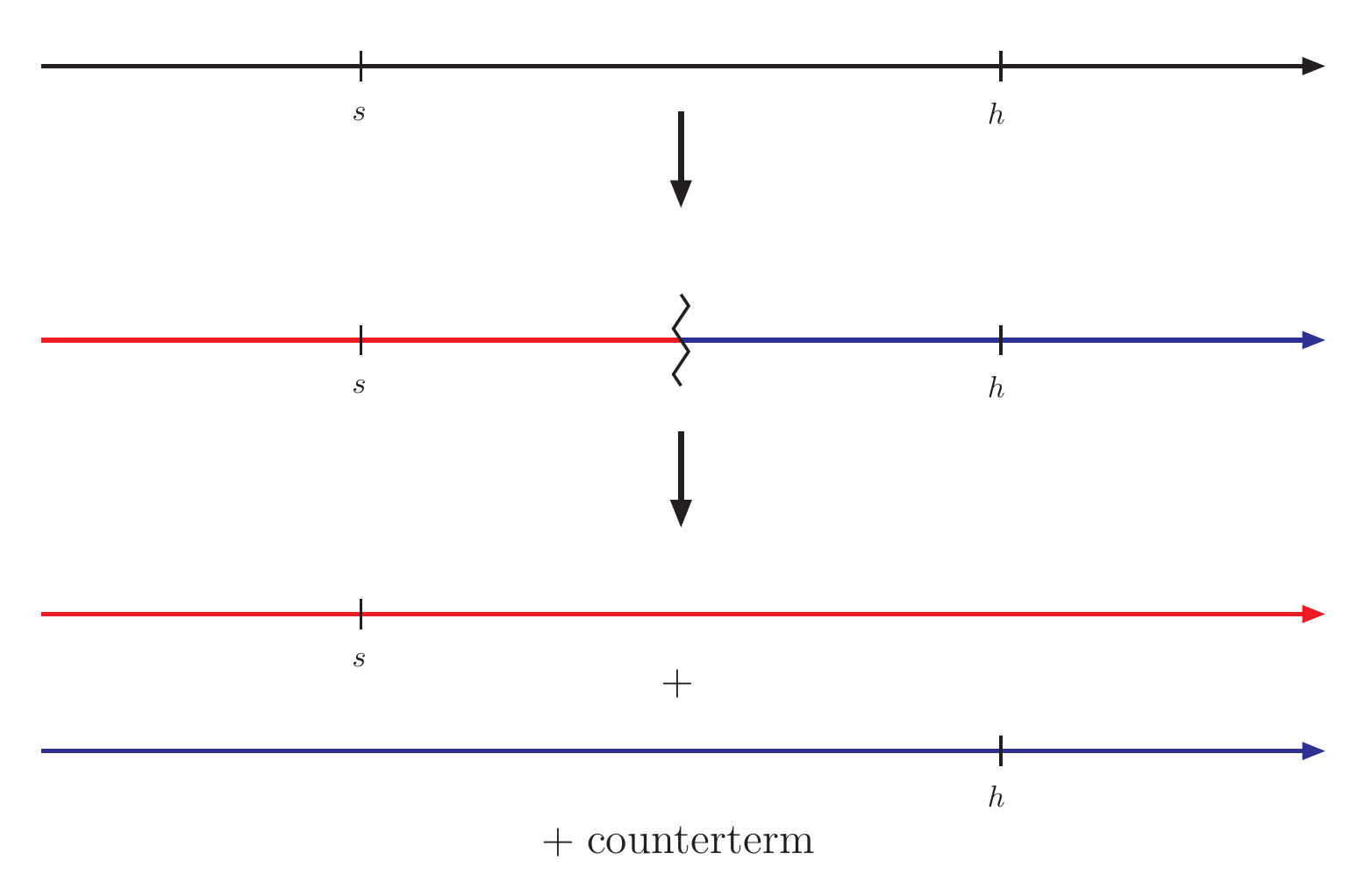}

\caption{\label{fig:decomposition}Pictorial representation of the decomposition
of loop integrals described in detail in appendix \ref{sec:Decomposition}.
A loop integral that involves both a soft $s$ and a hard scale $h$
can in the limit $s\ll h$ be divided at an arbitrary intermediate
scale and decomposed into two independent loop integrals and a counterterm
that cancels possible divergences introduced in this process. Each
of these two integrals involves only a single scale whereas the other
scale factors out of the corresponding integral. This allows a direct
power counting of the IR scaling behavior of the individual contributions.}

\end{figure}

\begin{figure}
\includegraphics[scale=0.75]{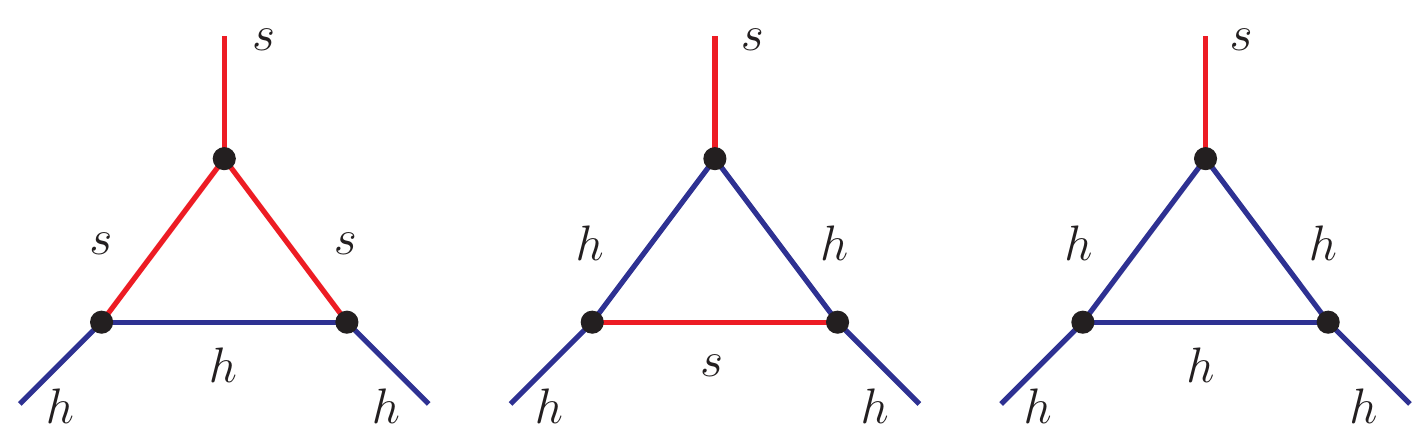}

\caption{IR singular kinematic regions in the decomposition of the triangle
integrals contributing to the 3-point functions in the soft particle
limit. There are two different kinematic regions that involve soft
loop momenta corresponding to the two inequivalent ways to route the
hard momentum through the loop.\label{fig:soft-loops}}

\end{figure}

Now let us discuss how to determine the IR behavior of the integrals
arising from the loop corrections in the DSEs. Inserting the parameterization
eq. (\ref{eq:general-scaling}) into a general 1-loop integral ${\cal G}$
yields a sum of different integrals of the form \begin{equation}
{\cal G}=\prod_{v}\sum_{l_{v}}\Lambda^{-2\left(\sum_{v}\left(\delta_{v}+\delta_{v}^{l_{v}}\right)\right)}\int\!\!\frac{d^{d}k}{\left(2\pi\right)^{d}}{\cal K}_{\{l_{v}\}}\left(\left\{ \left(k+Q\right)^{2}\right\} ,\left\{ Q^{2}\right\} \right)\prod_{m}\left(k+Q_{m}\right)^{2\left(\delta_{m}-1\right)}p_{l_{v}}^{2}\left(\left\{ \left(k+Q\right)^{2}\right\} ,\left\{ Q^{2}\right\} \right)^{\delta_{v}^{l_{v}}}\label{eq:general-integral}\end{equation}
where $m$ and $v$ run over the internal propagators and vertices
in the loop and $l_{v}$ over the the different IR sensitive limits
of the vertex $v$. The $Q_{i}$ are linear combinations of the external
momenta and the functions $\gamma_{v}^{l_{v}}$ as well as any other
analytic terms arising from the tensor decomposition are combined
in the kernels ${\cal K}_{\{l_{v}\}}$. They may involve additional
power law divergences in the external momenta alone but are analytic
in the loop momentum. Such integrals arise in addition to the explicitly
shown IR sensitive kinematical cases also for the individual terms
from the tensor decomposition eq. (\ref{eq:tensor-dec}). We note
in particular that due to the scaling form eq. (\ref{eq:general-scaling})
the dependence on the scale $\Lambda$ factors out of the actual integral.\\
The scalar integrals that remain after an appropriate tensor decomposition
are in general far too complicated to be performed analytically. However,
in the special case that the vertices are constant explicit solutions
are known. The 1-loop two-point integrals are given by the simple
analytic form

\begin{equation}
\int\frac{d^{d}k}{\left(2\pi\right)^{d}}\left(k^{2}\right)^{\nu_{1}}\left(\left(k-q\right)^{2}\right)^{\nu_{2}}=\left(4\pi\right)^{-\frac{d}{2}}\frac{\Gamma\left(\frac{d}{2}+\nu_{1}\right)\Gamma\left(\frac{d}{2}+\nu_{2}\right)\Gamma\left(-\frac{d}{2}-\nu_{1}-\nu_{2}\right)}{\Gamma\left(-\nu_{1}\right)\Gamma\left(-\nu_{2}\right)\Gamma\left(d+\nu_{1}+\nu_{2}\right)}\left(q^{2}\right)^{\frac{d}{2}+\nu_{1}+\nu_{2}}\end{equation}
for $d/2\!+\!\nu_{1}\!+\!\nu_{2}\leq0,$ $d/2+\nu_{1}\geq0$ and $d/2+\nu_{2}\geq0$,
where the expression is convergent. An analytic expression for the
corresponding IR 3-point integrals in terms of hypergeometric functions
is known \cite{Anastasiou:1999ui,Boos:1987bg} and we discuss how
to extend this result to the Euclidean regime of these integrals in
\cite{Alkofer:2008dt}. \\
However, in order to solely determine the IR scaling of a given
integral an explicit solution is not required. In the limit that some
external scales tend to zero and others stay finite one has a clear
scale separation. As is well known from effective field theories and
shown in detail in appendix \ref{sec:Decomposition} this allows one
to effectively decouple these scales. The basic steps are pictorially
illustrated in fig. \ref{fig:decomposition}. An integral involving
both hard and soft scales is first divided at an arbitrary intermediate
scale. The now limited support of the two integrals allows one to
Taylor-expand the integrand so that in the IR limit, where the lowest
order is exact, the individual integrals depend only on either soft
or hard scales and the respective other scales factor out of the integral.
Next, the two integrals are extended over the whole scale range again.
This leaves two correction terms that can be expanded once more and
combined to a scale-independent integral extending over all scales.
It merely presents a counterterm to cancel possible divergences introduced
in the process of separating the initial integral. Thereby the initial
integral is effectively decomposed into two separate integrals that
depend only on one of the two classes of scales respectively, whereas
in the soft part the dependence on the hard scales factors out of
the integral. On dimensional grounds such integrals must scale as
a function of this single scale and the power law exponents can then
be determined by a mere power counting analysis. This works since
the integrals are effectively dominated by the poles of the integrand
that are in turn determined by the external momenta. In particular,
these distinct integrals can each feature a different IR scaling behavior.
Thereby, when assessing the IR scaling of a given correlation function
via its DSE, all these partial integrals over the different kinematic
regions of the initial loop integral can be analyzed individually
and the most IR divergent term determines the IR scaling. As discussed
in appendix \ref{sec:Decomposition}, it can be necessary to repeat
this decomposition when multiple soft and hard scales are present.
After a complete decomposition the different kinematic contributions
of the initial integral can be displayed and treated just as distinct
Feynman graphs, as shown in fig. \ref{fig:soft-loops} for the soft-particle
limit of the 3-point function, where the momenta of the internal propagators
in the individual contributions are restricted to be soft $\left(s\right)$
respectively hard $\left(h\right)$ as indicated. 

\begin{figure}
\begin{minipage}[t]{0.5\columnwidth}%
\includegraphics[scale=0.75]{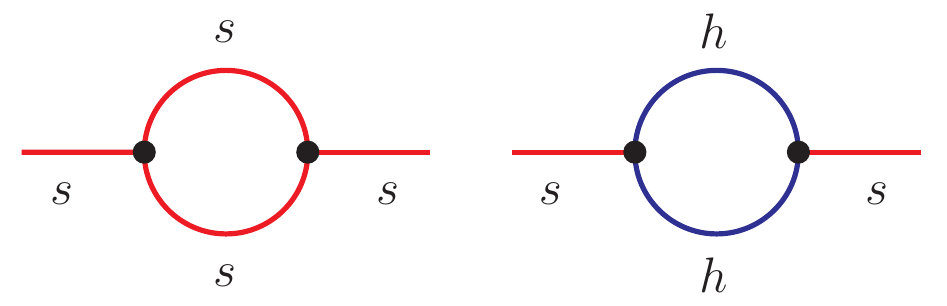}%
\end{minipage}%
\begin{minipage}[t]{0.5\columnwidth}%
\includegraphics[scale=0.75]{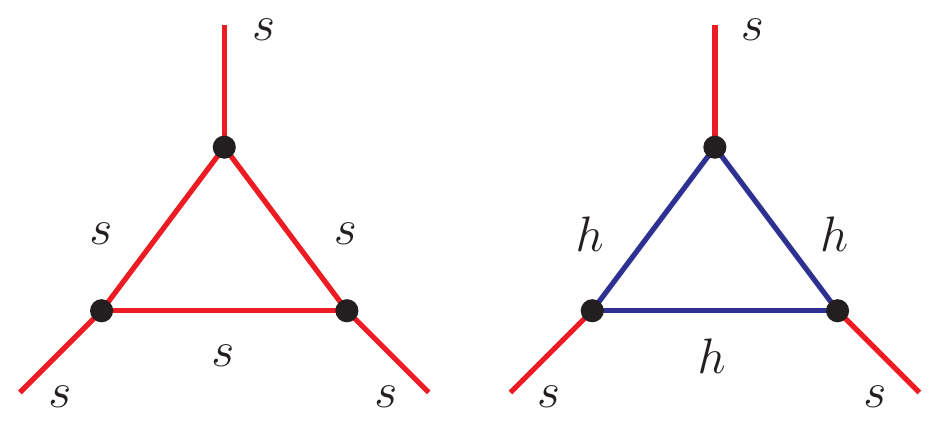}%
\end{minipage}

\caption{Possibly dominant kinematic regions of 1-loop integrals of the propagator
(left) and the 3-point function in the uniform limit (right). The
labels $s$ and $h$ denote soft momenta that vanish when the scaling
variable tends to zero and hard momenta that stay finite in the limit,
respectively.\label{fig:uni-loops}}

\end{figure}
Furthermore, it is necessary to consider the presence of mass scales
in the IR analysis. In addition to the case of theories that have
a sector which involves explicit mass scales at the classical level
which is coupled to a conformal sector, like QCD, there is even the
possibility of dynamical mass generation in theories that are scale-free
at the classical level, like Yang-Mills theory studied below. In general
a massive behavior of a (scalar) propagator $D$ is defined by

\begin{equation}
\lim_{p\to0}D\left(p^{2}\right)=\mathrm{const}\:.\end{equation}
This yields for the corresponding IR exponent $\delta$

\begin{equation}
D\left(p^{2}\right)=\frac{Z\left(p^{2}\right)}{p^{2}}\xrightarrow[p\to0]{}\frac{p^{2\delta}}{p^{2}}\sim p^{0}\quad\Rightarrow\quad\delta=1\:.\end{equation}
In this case the IR scaling is broken since an explicit mass scale

\begin{equation}
m^{2}\equiv\lim_{p\to0}D^{-1}\left(p^{2}\right)\end{equation}
is present and the propagator can be parameterized e.g. by the alternative
forms

\begin{equation}
D\left(p^{2}\right)=\frac{\tilde{Z}\left(p^{2}\right)}{p^{2}+m^{2}}=\frac{1}{p^{2}+M^{2}\left(p^{2}\right)}\quad\mathrm{with}\quad\tilde{Z}\left(p^{2}\right)\equiv\frac{p^{2}+m^{2}}{p^{2}}Z\left(p^{2}\right)\;,\quad M\left(p^{2}\right)\equiv p^{2}\left(\frac{1}{Z\left(p^{2}\right)}-1\right)\:.\end{equation}
In this form it is clear that integrals can also be dominated by
scales of the order of the mass, as discussed in detail in appendix
\ref{sec:Decomposition}. In the presence of mass scales there are
relevant contributions from corresponding hard momentum scales in
the loop integration even in the case that all external scales are
soft, as shown in fig. \ref{fig:uni-loops}. If these hard contributions
dominate such a dynamically generated mass presents a self-consistent
solution of the DSE system and poses an additional possibility for
the IR behavior that has to be taken into account. \\
 In the uniform limit the scaling is very different in the massive
and massless case. When there are only massive propagators in a loop
integral the integral is entirely dominated by hard scales even in
the uniform limit and does not scale with the external soft scales,
see appendix \ref{sec:Decomposition}. When only some propagators
are massive the integral can again be decomposed and receives contributions
from both soft and hard momentum regions, as is again shown in appendix
\ref{sec:Decomposition} and is depicted in fig. \ref{fig:uni-loops}.
In contrast, in the massless case the integrals are IR dominated in
the uniform limit and the contribution from hard loop momenta is strongly
suppressed. This can be seen by expanding the analytic functions ${\cal K}_{\{l_{v}\}}$
in eq. (\ref{eq:general-integral}) as well as the arising propagators
in the limit $q_{i}\ll k$, where $k$ is the loop momentum,

\begin{align}
\frac{1}{\left(k+q_{i}\right)^{2\left(1-\alpha\right)}} & =\frac{1}{k^{2\left(1-\alpha\right)}}\left(1+2\frac{q_{i}\cdot k}{k^{2}}+\frac{q_{i}^{2}}{k^{2}}\right)^{\alpha-1}\label{eq:hard-suppression}\\
 & \approx\frac{1}{k^{2\left(1-\alpha\right)}}\left(1+2\left(\alpha-1\right)\frac{q_{i}\cdot k}{k^{2}}+\left(\alpha-1\right)\frac{q_{i}^{2}}{k^{2}}+2\left(\alpha-1\right)\left(\alpha-2\right)\frac{\left(q_{i}\cdot k\right)^{2}}{k^{4}}+O\left(\frac{q_{i}^{3}}{k^{3}}\right)\right)\nonumber \end{align}
The leading order term is indeed independent of the external momenta.
In case there are no other scales involved, i.e. in particular $\alpha\neq1$,
cf. appendix \ref{sec:Decomposition}, this scale independent integral
as well as the third term is canceled or removed in the renormalization
process, as discussed below. The linear term vanishes likewise in
the symmetric integration so that only the last term remains and the
contribution from hard modes is actually suppressed in $q_{i}^{2}/k^{2}$.
Yet, when there is any explicit scale in the integral, like a hard
external momentum or a mass (corresponding to $\alpha=1$), the leading
term in the expansion eq. (\ref{eq:hard-suppression}) is not canceled
and there is no suppression. In order to treat the different cases
within a common framework and allow for all possible solutions, including
those where masses are dynamically generated, we count in all cases
the contributions from both the hard and soft momentum region of the
loop integral.\\
In case a mass is generated dynamically it arises from quantum
fluctuations in the \textquotedbl{}mid-momentum regime\textquotedbl{},
i.e. hard scales of the order of the generated mass. To establish
whether a massive solution actually exists and to obtain a quantitative
value for the mass requires a numerical solution of the functional
equations for all momenta. However, a mere IR analysis can nevertheless
determine if a dynamically generated mass presents a possible dynamical
solution of the DSE system. This is the case since the above expansion
eq. (\ref{eq:hard-suppression}) shows that the different cases of
a massless and a massive propagator are effectively discriminated
by their IR exponents. In the power counting study below we implement
this qualitatively different scaling of the propagators arising in
loop integrals via the symbol $\mu_{i}$ where the index $i$ stands
for the corresponding particle species and which is defined by

\begin{equation}
\mu_{i}=\left\{ \begin{array}{cccc}
0 & \mathrm{for} & \delta_{i}=1 & \mathrm{,\; i.e.\: a\: massive\: IR\: behavior}\\
1 & \mathrm{for} & \delta_{i}\neq1\end{array}\right.\label{eq:mass-counting}\end{equation}
Since eq. (\ref{eq:hard-suppression}) shows that the contribution
from the hard momentum regime of a loop integral is suppressed only
if all propagators are massless, the suppression of the whole loop
integral is then given by adding the product of the $\mu_{i}$ of
all propagators in the loop to the loop's IR exponent. We stress that
although in the considered region of hard momenta the dressing functions
do not feature the IR scaling form, the above prescription is meaningful
for the power counting analysis since the massive behavior is nevertheless
reflected in the anomalous IR power laws by the IR exponent $\delta_{i}=1$.\\
The loop integrals in the DSEs are UV divergent and must be renormalized.
Since the leading Green functions have positive mass dimension power
law divergences seem to pose a threat to a naive power counting analysis
because hard momenta could contribute even in the case that all external
momenta are soft. Actually, the above suppression explicitly shows
the known fact that all power law divergences, like e.g. the quadratic
divergences in the propagators, are momentum independent in scale
invariant theories. If present, these identically cancel between the
different loop graphs in a DSE. Whereas this is generally guaranteed
by the perturbative UV behavior so that the contributions from asymptotically
large momenta $k\gg\Lambda$ to the integral cancel, in the IR scaling
limit the cancellation of contributions from hard modes is according
to eq. (\ref{eq:hard-suppression}) already realized for loop momenta
$k\gg p$. Similarly there are dedicated counterterms that by construction
explicitly cancel any additional singularities introduced in the above
decomposition, cf. appendix \ref{sec:Decomposition}. This leaves
at most logarithmic momentum-dependent divergences to be removed within
the renormalization procedure. As shown e.g. in \cite{vonSmekal:1997is,Lerche:2002ep}
this is possible multiplicatively by corresponding wave-function and
vertex renormalization factors that multiply the corresponding tree
level terms as well as the loop corrections in the DSEs. Since the
logarithmic momentum-dependence is not considered within the power
counting analysis anyhow, it is clear that the problem of renormalization
does not interfere with the IR scaling analysis as could have been
expected considering that they operate in opposite limits of the loop
integrals. In principle the analysis can also be performed directly
for already subtracted DSEs where the issue is omitted from the outset.
Thereby, in general UV divergences and the detailed renormalization
procedure do not affect the IR analysis. Yet, there is one important
exception to this since under certain conditions discussed below the
renormalization prescription can determine the boundary conditions
for the DSE system \cite{Fischer:2008uz}.\\
After a complete decomposition of the loop integrals in the DSEs,
the IR scaling of each contribution is determined via power counting
of the anomalous and canonical dimensions of the involved Green functions.
The most divergent contribution - corresponding to the minimal IR
exponent - dominates and determines the scaling of the Green function
on the left hand side of the DSE. Finally it is important to note
that there can be various cancelations caused by the symmetries of
the considered theory that require a careful analysis. This will be
discussed in detail in section \ref{sec:Soft-singularities} for the
case of Landau gauge Yang-Mills theory.

\section{Disproval of the naive infrared slavery scenario\label{sec:IR-slavery}}

After this general section we will now begin our IR analysis of Yang-Mills
theory. The equations for the propagators of Landau gauge Yang-Mills
are given in figs. \ref{fig:ghost-DSE} and \ref{fig:gluon-DSE}.
These equations have been studied extensively with appropriate ansaetze
for the vertices. The corresponding DSEs for the primitively divergent
vertex functions of Yang-Mills theory are given in figs. \ref{fig:ghost-gluon-DSE-1}
to \ref{fig:4-gluon-DSE}. For the ghost-gluon vertex there are two
qualitatively different versions derived via the ghost- respectively
gluon-part of the path integral. The leading order in a skeleton expansion
is given by the two triangle diagrams that have been analyzed in \cite{Schleifenbaum:2004id}
within a semi-perturbative analysis, where it was found that this
vertex is hardly changed from its tree-level form. The other vertices
have so far been discussed only via IR scaling analyses \cite{Alkofer:2004it,Fischer:2006vf,Huber:2007kc}
which we will detail in this work. Explicit IR results for the $3$-point
vertices are given in \cite{Alkofer:2008dt}. \\
Before we study the implications of the system of equations for
the IR limit of Yang-Mills theory, we will demonstrate the power counting
method explicitly for an important example. Consider the first two
diagrams of the second line in the DSE for the 3-gluon vertex fig.
\ref{fig:3-gluon-DSE}. The contribution from these two graphs to
the right hand side of the DSE is

\begin{align}
\Delta\Gamma_{\mu\nu\rho}^{abc}\left(q_{1},q_{2}\right) & =\tfrac{1}{2}\int\frac{d^{4}k}{\left(2\pi\right)^{4}}\left(\left(\Gamma_{0}\right)_{\rho\mu\alpha\beta}^{cade}D_{\alpha\gamma}^{df}\left(k-q_{2}\right)\Gamma_{\gamma\delta\nu}^{fgb}\left(k-q_{2},-k,q_{2}\right)D_{\delta\beta}^{ge}\left(k\right)\right.\nonumber \\
 & \qquad\qquad\qquad\left.+\left(\Gamma_{0}\right)_{\nu\rho\alpha\beta}^{bcde}D_{\alpha\gamma}^{df}\left(k-q_{1}\right)\Gamma_{\gamma\delta\mu}^{fga}\left(k-q_{1},-k,q_{1}\right)D_{\delta\beta}^{ge}\left(k\right)\right)\:,\end{align}
where $D$ represents the dressed gluon propagator, $\Gamma$ the
dressed vertices distinguished by their indices, and $\Gamma_{0}$
the bare versions. We note that each of the two integrals depends
on only one of the two independent external momenta, because the bare
4-gluon vertex is momentum independent. Therefore, the two integrals
cannot exactly cancel each other for general momenta. With the uniform
IR scaling exponents $\delta_{gl}$ for gluon propagators and the
corresponding uniform exponent $\delta_{3g}^{u}$ for the 3-gluon
vertex each of the integrals as well as its sum scales in the uniform
limit as

\begin{equation}
\Delta\Gamma_{\mu\nu\rho}^{abc}\left(q_{1},q_{2}\right)\xrightarrow[p^{2}\rightarrow0]{}p^{4}\left(\left(p^{2}\right)^{-1+\delta_{gl}}\right)^{2}\left(p^{2}\right)^{\frac{1}{2}+\delta_{3g}^{u}}\sim\left(p^{2}\right)^{\frac{1}{2}+\delta_{3g}^{u}+2\delta_{gl}}\:.\end{equation}
We see that the canonical scaling of the $3$-gluon vertex given by
the $\frac{1}{2}$ drops out since it appears both on the left and
right hand side of the DSE and as expected it is sufficient to count
only anomalous IR exponents in the uniform scaling limit. Each of
the contributions on the right hand side of the DSE - or several of
them - could dominate and determine the scaling of the 3-gluon vertex
on the left hand side of the DSE. The leading term is the one with
the smallest IR exponent. Correspondingly, the IR exponent of the
two exemplary graphs has to be larger or equal to the left hand side
in case they dominate. This leads to the very general condition

\begin{equation}
\delta_{3g}^{u}\leq\delta_{3g}^{u}+2\delta_{gl}\quad\Rightarrow\quad\delta_{gl}\geq0\:.\end{equation}
The only other possibility would be that the two integrals are canceled
identically by other diagrams in the DSE which correspondingly would
have to have precisely the same kinematic dependence. We pointed out
before that the two graphs have a very special kinematic structure
given by

\begin{equation}
\Delta\Gamma_{\mu\nu\rho}^{abc}\left(q_{1},q_{2}\right)=F_{\mu\nu\rho}^{abc}\left(q_{1}\right)+G_{\mu\nu\rho}^{abc}\left(q_{2}\right)\:.\end{equation}
Evidently all other graphs involve the two momenta in a manifestly
non-linear way - e.g. already the propagators in the loop induce manifest
non-linearities - and thereby cannot have the above simple property.
For instance consider the case of the third gluon loop graph that
includes the proper 4-gluon vertex. Its integral representation is

\begin{equation}
\tfrac{1}{2}\!\int\!\!\frac{d^{d}k}{\left(2\pi\right)^{d}}\left(\Gamma_{0}\right)_{\mu\delta\alpha}^{aed}\left(\! q_{1},k,-\! k\!-\! q_{1}\!\right)\frac{Z\left(k\!+\! q_{1}\right)}{\left(k\!+\! q_{1}\right)^{2}}\!\left(\!\delta_{\alpha\beta}\!-\!\frac{\left(k\!+\! q_{1}\right)_{\alpha}\left(k\!+\! q_{1}\right)_{\beta}}{\left(k\!+\! q_{1}\right)^{2}}\!\right)\!\Gamma_{\nu\rho\beta\gamma}^{bcde}\left(\! q_{2},-\! q_{1}\!-\! q_{2},k\!+\! q_{1},-\! k\!\right)\frac{Z\left(k\right)}{k^{2}}\!\left(\!\delta_{\gamma\delta}\!-\!\frac{k_{\gamma}k_{\delta}}{k^{2}}\!\right)\end{equation}
where the gluon dressing functions $Z$ generally involve non-integer
powers. Even in the simplest case $\delta_{gl}=1$ where the propagators
become trivial the above property would be in contrast to the 1PI
nature of the proper $4$-point vertex. Since the momentum dependence
of the other contributions in the DSE is even more non-linear it is
fair to conclude that there are no identical cancelations between
the individual diagrams in fig. \ref{fig:3-gluon-DSE}. This yields
the direct constraint that the gluon dressing function \emph{cannot}
be singular as found in the Mandelstam approximation \cite{Mandelstam:1979xd}.
The same relation is obtained from the gluon-loop corrections given
by the last three diagrams in the second line of the 4-gluon vertex
DSE fig. \ref{fig:4-gluon-DSE}. Note that this result is a direct
prediction of the full vertex DSEs and does not involve any assumption
or approximation. Moreover, it is independent of the detailed renormalization
prescription used for the DSEs. Interestingly, all studies that found
an IR enhanced gluon propagator made uncontrolled assumptions on the
form of the 3-gluon vertex, whereas such a non-enhanced IR behavior
of the propagator was indeed found as soon as the vertex equations
were considered dynamically within simplified semi-perturbative analyses,
cf. \cite{Roberts:1994dr}.

\section{Conformal scaling\label{sec:Conformal-scaling}}

\begin{figure}
\includegraphics[scale=0.5]{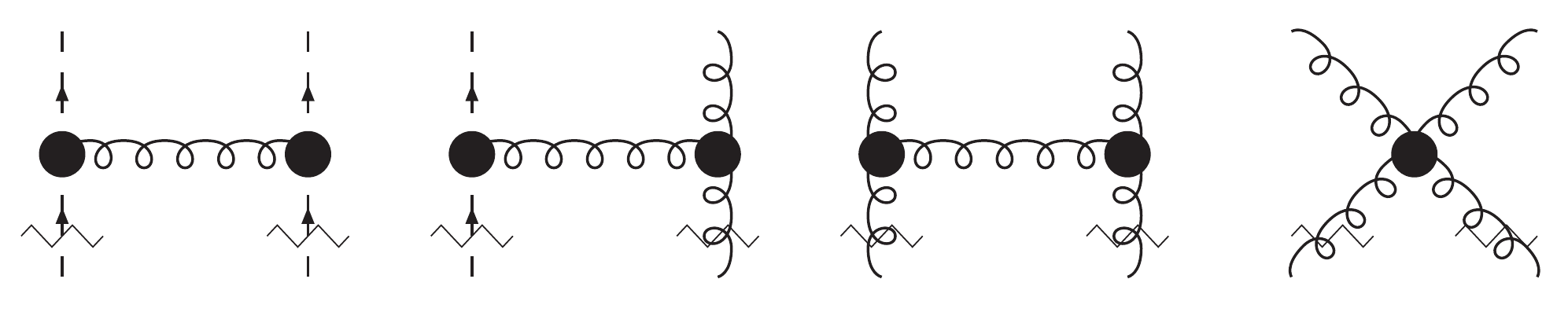}

\caption{\label{fig:sekelton}Possible extensions of given graphs to generate
higher orders in the skeleton expansions. The crossed out propagators
are part of the initial graph and are not counted. }

\end{figure}

In this section we will perform an IR fixed point analysis for Landau
gauge Yang-Mills theory but will first neglect the possibility of
kinematic singularities or dynamically generated mass scales. Thereby
we consider the idealized, conformal case that all external momenta
as well as the considered correlation functions scale with a single
uniform scaling variable $p_{u}$, as defined in eq. (\ref{eq:uniform-variable}),
and are finite in other kinematic limits. In this case eq. (\ref{eq:general-scaling})
reduces to \begin{equation}
\Gamma_{v}\left(q_{1}^{2},q_{1}\!\cdot\! q_{2},\cdots,q_{n}^{2}\right)\approx\gamma_{v}^{u}\left(q_{1}^{2},q_{1}\!\cdot\! q_{2},\cdots,q_{n}^{2}\right)\cdot\left(\frac{p_{u}^{2}\left(q_{1}^{2},\cdots,q_{n}^{2}\right)}{\Lambda_{QCD}^{2}}\right)^{\delta_{v}^{u}}\:.\label{eq:uniform-scaling}\end{equation}
Correspondingly, we neglect the possibility that contributions from
hard loop momenta are relevant in the IR limit and assume that only
the IR regime dominates all loop integrals. As will be shown in the
next section this assumption is too simplified and the solution structure
obtained in this section is incomplete.\\
Since the Dyson-Schwinger equations form an infinitely coupled
system of equations it is necessary to reduce it to a manageable form,
as discussed in sec. \ref{sec:IR-analysis}. This is done via a skeleton
expansion that presents a loop expansion in terms of dressed primitively
divergent correlation functions. The skeleton expansion can be generated
from the leading order graphs by a finite set of extensions that increase
their loop order \cite{Alkofer:2004it} as shown in fig. \ref{fig:sekelton}.
As a necessary condition for the skeleton expansion these extensions
must not increase the IR exponents for a given Green function since
otherwise successive extensions would make it arbitrary singular.
Thereby, assuming that the skeleton expansion is not explicitly divergent,
provides additional constraints for the IR exponents of the primitively
divergent vertices. \\
When there is only a single uniform scaling variable $p_{u}$,
so that all external momenta scale with it, $q_{i}^{2}\sim p_{u}^{2}$,
up to possible subleading corrections an integral has to scale as
a power of it (or possibly as a logarithm in case the naive power
counting yields a constant). As noted before, in four dimensions
all canonical momentum dependence cancels and it turns out that it
suffices to count anomalous powers of $p_{u}^{2}$ to assess the IR
behavior of a general loop correction. The leading dynamical contribution
on the right hand side of its DSE determines the scaling of a given
Green function. With the corresponding uniform IR exponents for the
ghost propagator $\delta_{gh}$, the gluon propagator $\delta_{gl}$,
the ghost-gluon vertex $\delta_{gg}$, the 3-gluon vertex $\delta_{3g}$
and the 4-gluon vertex $\delta_{4g}$ we can analyze the IR scaling
limit of the DSE system for the five primitively divergent Green functions.
Here we skip the superscript $u$ on the uniform exponents as it appears
in eq. (\ref{eq:uniform-scaling}) for better readability in this
section. From figs. \ref{fig:ghost-DSE} and \ref{fig:gluon-DSE}
one reads off the power counting relations for the IR exponents of
the propagators%
\footnote{The tadpole contribution in the gluon DSE does not appear here, since
it does not depend on the external momentum.%
}

\begin{align}
-\delta_{gh} & =\min\left(0,\delta_{gg}+\delta_{gh}+\delta_{gl}\right)\:,\nonumber \\
-\delta_{gl} & =\min\left(0,\delta_{3g}+2\delta_{gl},\delta_{gg}+2\delta_{gh},2\delta_{3g}+4\delta_{gl},\delta_{4g}+3\delta_{gl}\right)\label{eq:uniform-prop-rel}\end{align}
and correspondingly for the vertex functions from figs. \ref{fig:ghost-gluon-DSE-1}
to \ref{fig:4-gluon-DSE}

\begin{align}
\delta_{gg} & =\min\left(0,2\delta_{gg}+2\delta_{gh}+\delta_{gl},\delta_{3g}+\delta_{gg}+\delta_{gh}+2\delta_{gl}\right)\:,\label{eq:uniform-ver-rel}\\
\delta_{gg} & =\min\left(0,2\delta_{gg}+2\delta_{gh}+\delta_{gl},2\delta_{gg}+\delta_{gh}+2\delta_{gl},3\delta_{gg}+2\delta_{gh}+3\delta_{gl},\delta_{3g}+2\delta_{gg}+\delta_{gh}+4\delta_{gl}\right)\:,\nonumber \\
\delta_{3g} & =\min\left(0,2\delta_{gg}+3\delta_{gh},2\delta_{3g}+3\delta_{gl},\delta_{3g}+2\delta_{gl},\delta_{4g}+2\delta_{gl},3\delta_{3g}+5\delta_{gl},\delta_{4g}+\delta_{3g}+4\delta_{gl}\right)\:,\nonumber \\
\delta_{4g} & =\min\left(0,3\delta_{gg}+4\delta_{gh},3\delta_{3g}+4\delta_{gl},\delta_{4g}+2\delta_{gl},2\delta_{3g}+3\delta_{gl},\delta_{4g}+\delta_{3g}+3\delta_{gl},4\delta_{3g}+6\delta_{gl},\delta_{4g}+2\delta_{3g}+5\delta_{gl},2\delta_{4g}+4\delta_{gl}\right)\:.\nonumber \end{align}
Besides the constraint $\delta_{gl}\geq0$ obtained before there
are additional analogous constraints from the linear terms in the
vertex equations (\ref{eq:uniform-ver-rel}). These constraints are
weaker than those obtained from the corresponding RG equations studied
in \cite{Fischer:2006vf,Huber:2009wh} and are not sufficient to
ensure a unique solution of the system of equations. Therefore, we
will not exploit them in the following and thereby circumvent the
problem of possible cancelations discussed above. 

In its present form the nonlinear system of the DSE conditions (\ref{eq:uniform-prop-rel})
and (\ref{eq:uniform-ver-rel}) is rather complicated due to the involved
minimum functions. However, if one requires that there should exist
a stable skeleton expansion, the extension graphs in fig. \ref{fig:sekelton}
yield the additional much stronger constraints

\begin{align}
2\delta_{gg}+2\delta_{gh}+\delta_{gl} & \geq0\:,\nonumber \\
\delta_{3g}+\delta_{gg}+\delta_{gh}+2\delta_{gl} & \geq0\:,\nonumber \\
2\delta_{3g}+3\delta_{gl} & \geq0\:,\nonumber \\
\delta_{4g}+2\delta_{gl} & \geq0\:.\label{eq:se-rel}\end{align}
The combination of IR exponents in the first two of these constraints
is precisely the one arising from the two triangle diagrams in the
first equation for the exponent of the ghost-gluon vertex eq. (\ref{eq:uniform-ver-rel}),
so that this equation becomes unique whereas the second one remains
non-trivial

\begin{equation}
\delta_{gg}=0\;\wedge\;\delta_{gg}=\min\left(0,2\delta_{gg}+2\delta_{gl}+\delta_{gh}\right)\:.\end{equation}
Since both of these equations have to hold simultaneously it is clear
that the second equation has the trivial solution, too. To see that
it is not important which DSE we start with, consider only the second
equation and assume that the second term dominates. In this case the
ghost-gluon vertex would be given by $\delta_{gg}=-2\delta_{gl}-\delta_{gh}$.
Plugging this in the first constraint from the skeleton expansion,
eq. (\ref{eq:se-rel}), yields $\delta_{gl}\leq0$. Since we know
from section \ref{sec:IR-slavery} that $\delta_{gl}$ is non-negative,
only the trivial solution $\delta_{gl}=\delta_{gh}$ remains.\\
This non-renormalization condition has previously been used as
a starting point in the analysis \cite{Alkofer:2004it}. It was supported
by the gluon transversality in Landau gauge but required the additional
assumption that the ghost-gluon scattering kernel should not be strongly
divergent. Here it arises directly from the physical requirement of
a stable skeleton expansion. The two other constraints in (\ref{eq:se-rel})
precisely remove the non-linearities in the equations (\ref{eq:uniform-ver-rel})
for the gluon vertices which could lead to a self-consistent enhancement
of these equations. Inserting this IR scale independence of the ghost-gluon
vertex into the other equations and using the constraints eqs. (\ref{eq:se-rel})
as well as the previously shown relation $\delta_{gl}\geq0$ yields

\begin{align}
-\delta_{gh} & =\min\left(0,\delta_{gh}+\delta_{gl}\right)\:,\nonumber \\
-\delta_{gl} & =\min\left(0,2\delta_{gh}\right)\:,\nonumber \\
\delta_{3g} & =\min\left(0,3\delta_{gh}\right)\:,\nonumber \\
\delta_{4g} & =\min\left(0,4\delta_{gh},3\delta_{3g}+4\delta_{gl}\right)\:.\label{eq:red-uniform-ver-rel}\end{align}
As it stands the naive system of equations above has only the trivial
solution that all anomalous IR exponents vanish since for $\delta_{gl}\geq0$
the first equation yields directly $\delta_{gh}=0$ and the rest follows
trivially. \\
However, so far it has not been taken into account in our scaling
analysis that the DSEs have to be renormalized. As shown in \cite{vonSmekal:1997is,Lerche:2002ep}
when a propagator is divergent, it is possible to do this renormalization
at $Q^{2}=0$ which cancels the tree level part identically. Thereby,
the renormalization condition can act as a boundary condition for
the DSEs, as observed in \cite{Fischer:2008uz}. Since the exponent
of the gluon propagator is non-negative, as shown before, this renormalization
prescription is not possible there. In contrast due to this positivity
the mixed loop correction in the ghost equation proves that the ghost
propagator can indeed be divergent and the two exponents are then
connected by $\delta_{gh}=-\delta_{gl}/2$. Because of this connection
the equation for the gluon propagator is dominated by the ghost loop,
but it becomes trivial and does not determine $\delta_{gl}$. The
solution depends therefore on a free parameter $\delta_{gl}\equiv2\kappa\geq0$,
$\delta_{gh}=-\kappa\leq0$. With the expressions for the propagators
this immediately gives $\delta_{3g}=-3\kappa$ from the ghost loop
which shows that the ghost loop dominates the gluon loop also in the
four gluon vertex DSE and we have $\delta_{4g}=-4\kappa$. Therefore,
there is only a single\emph{ scaling solution} of this system depending
on a real parameter $\kappa\geq0$ when the relations arising from
the condition of a stable skeleton expansion eq. (\ref{eq:se-rel})
are taken into account. This is the solution obtained previously in
\cite{Alkofer:2004it,Fischer:2006vf,Huber:2007kc} where it was shown
that the above ghost dominance mechanism holds for arbitrary $n$-point
functions. Ghost loop contributions to gluonic correlation functions
and minimal mixed loop contributions to correlation functions involving
ghosts dominate and all gluonic corrections are suppressed due to
the scaling of the propagators despite the strongly divergent gluonic
vertex functions. Here, in particular the trivial solution $\delta_{i}=0\:,\;\forall i$,
obtained for a generic renormalization prescription is also contained
in the above scaling solution for $\kappa=0$. The restriction to
solutions of the scaling type arises here since we implicitly assumed
that the integrals are always dominated by modes in the vicinity of
soft external momentum scales. In the next section we will see that
this is too simplified and that the result can be changed by the presence
of large  scales.

\section{Inclusion of soft singularities and masses\label{sec:Soft-singularities}}

As discussed in detail in section \ref{sec:IR-analysis}, the IR counting
is complicated by the possibility of kinematic divergences or dynamical
mass generation. In these cases there are hard external scales present
that do not tend to zero when taking the IR limit and care has to
be taken to assess how a certain correlation function scales with
the soft momenta. Whereas for the propagators there are unique anomalous
IR exponents, in general there can be different IR exponents that
describe how a correlation function scales in different kinematic
sections. The uniform limit where all external scales go to zero uniformly
presents the conformal case studied above. Beyond this, there are
for the 3-point functions possibly distinct IR exponents when only
a single momentum vanishes, as illustrated in fig. \ref{fig:3-point-kinematics}.
According to eq. (\ref{eq:general-scaling}), we discriminate these
different kinematic limits by upper indices and denote the corresponding
uniform indices in the following by $\delta_{gg}^{u}$ and $\delta_{3g}^{u}$.
There are in addition the exponents for the ghost-gluon and 3-gluon
vertices, when a gluon momentum vanishes, $\delta_{gg}^{gl}$ and
$\delta_{3g}^{gl}$, respectively, and for the ghost-gluon vertex,
when a ghost momentum vanishes, $\delta_{gg}^{gh}$. As illustrated
in fig. \ref{fig:4-point-kinematics}, for the 4-gluon vertex, studied
in the uniform limit in \cite{Kellermann:2008iw}, there can be two
distinct additional kinematic exponents $\delta_{4g}^{gl}$ when one
or $\delta_{4g}^{2g}$ when two gluon momenta vanish. In addition
there could even be IR divergences when all external momenta are large
but differences of momenta become small. Because of the symmetry of
the vertex this can be described by one additional IR exponent $\delta_{4g}^{i}$
corresponding to the exchange of a soft momentum in an intermediate
channel.\\
Although possible, it is rather cumbersome to determine the IR
exponent of the 2-loop graphs appearing in some DSEs taking into account
the various possible kinematic singularities. Fortunately, from what
we know from the analysis in the uniform limit this should not be
necessary. There the observed strong ghost dominance strongly suppressed
gluonic contributions compared to the leading ghost loops. Because
of the absence of primitively divergent 4-point interactions involving
ghosts all of these leading contributions involve only 3-point vertices
and are 1-loop graphs. \emph{Assuming} that the additional kinematic
singularities do not entirely change this property motivates a truncation
scheme involving only dressed 3-point vertices neglecting all 2-loop
graphs and those involving dressed 4-point vertices. The only graphs
that involve bare 4-point vertices that are not 2-loop are those discussed
in sec. \ref{sec:IR-slavery} and which provided the mere constraint
$\delta_{gl}\geq0$. Therefore we will not have to discuss them here
again. The truncation we study here is also similar to the one obtained
from a 3-loop expansion of a 3PI action analyzed in \cite{Berges:2004pu}
but with the difference that in the DSEs there is one bare vertex
in every graph.\\
To recapitulate, the general assumptions of the IR analysis introduced
in section \ref{sec:IR-analysis} are

\begin{itemize}  \item[(i)] that the local degrees of freedom are valid to describe the system,  \item[(ii)] that the IR regime of the DSE system can be analyzed using a skeleton expansion and   \item[(iii)] that the IR is described by power law scaling. \end{itemize} Furthermore,
in this section only 

\begin{itemize} \item[(iv)] we truncate the system to 1-loop graphs to simplify the analysis. \end{itemize} We
emphasize that once solutions are found it can be explicitly verified
that they present consistent IR solutions of the \emph{full} DSE system.
Green functions are then dominated by the leading order skeleton contributions
involving primitively divergent Green functions, i.e. the assumptions
(i) \& (iv) are indeed fulfilled. The assumptions (ii) \& (iii) merely
limit the class of solutions that can be found with our technique.\\
In the case that several external scales are present the loop integral
can receive relevant contributions from fluctuations in the vicinity
of all these different scales. Therefore, as discussed in section
\ref{sec:IR-analysis}, it is necessary to decompose the momentum
integral into distinct integrals which in the IR limit are entirely
dominated in the respective soft or hard momentum regions. These
integrals involve only one characteristic scale each and their IR
scaling can then again be determined by power counting. Furthermore,
we have to take into account the possibility of dynamical mass generation.
Generally the presence of masses is similar to hard momentum scales
in that the hard momentum region of loop integrals can be important.
The extended power counting analysis that takes into account all these
aspects has been presented in section \ref{sec:IR-analysis} where
the possibility of dynamical mass generation was linked to the IR
exponents of the propagators via the auxiliary symbols in eq. (\ref{eq:mass-counting})
for the gluons $\mu_{gl}$ and ghosts $\mu_{gh}$.\\
In the presence of several different scales a mere power counting
of anomalous IR exponents is not sufficient anymore but the canonical
scaling of the integrals, propagators and vertices has to be considered.
In particular it is possible that tensor structures of a vertex involve
hard momenta and do not scale with the soft momentum. Correspondingly,
it is also necessary to discriminate between bare and dressed vertices
in this context. Therefore we will first assess the canonical scaling
of the appearing vertices in detail. Let us start with the bare ghost-gluon
vertex which in our conventions depends only on the outgoing ghost
momentum. If this momentum is soft the canonical scaling has to be
taken into account independent of the size of the other momenta and
vice versa. In addition, when the vertex is connected to an internal
gluon propagator the tensor in the direction of the gluon momentum
is canceled due to the transversality of the gluon propagator in Landau
gauge

\begin{equation}
q^{\mu}D_{\mu\nu}\left(p\right)=\left(q-\frac{p\cdot q}{p^{2}}p\right)^{\mu}D_{\mu\nu}\left(p\right)\equiv q_{\perp}^{\mu}D_{\mu\nu}\left(p\right)\:.\label{eq:transversality}\end{equation}
The dressed ghost-gluon vertex has two independent tensor structures
that can be chosen as arbitrary linear combinations of the 3 external
momenta. If only one external momentum is soft there is a tensor structure
that depends on hard momenta which will dominate as long as there
are no cancelations. Therefore a dressed vertex has generically no
canonical scaling whenever there are hard scales involved. However,
because of the above transversality this can be changed when there
is only one hard external scale - as is the case in propagator integrals
- but not if there are two independent ones.\\
The 3-gluon vertex has many tensor structures but it turns out
to be sufficient to analyze the tree level tensor

\begin{equation}
\left(\Gamma_{0}\right)_{\mu\nu\rho}^{abc}\left(p,q,r\right)=-igf^{abc}\left(\left(p-q\right)_{\rho}\delta_{\mu\nu}+\left(q-r\right)_{\mu}\delta_{\nu\rho}+\left(r-p\right)_{\nu}\delta_{\rho\mu}\right)\:.\end{equation}
When only one of the momenta is soft the tensor is in general of the
order of the hard momenta that dominate soft contributions. Other
possible tensor structures that depend only on soft momenta are likewise
subleading compared to the tree level tensor and the canonical scaling
is again not present. Gluon transversality cannot directly change
this here since the above tensor structure involves the metric tensor
which couples the two attached gluon propagators in a loop directly.
However, there are cases where the Bose symmetry of the vertex can
lead to additional cancelations as discussed below.\\
Let us now apply this to the individual diagrams to obtain the
corresponding equations for the IR exponents. The decomposition of
the full ghost DSE reads

\begin{center}\includegraphics[scale=0.7]{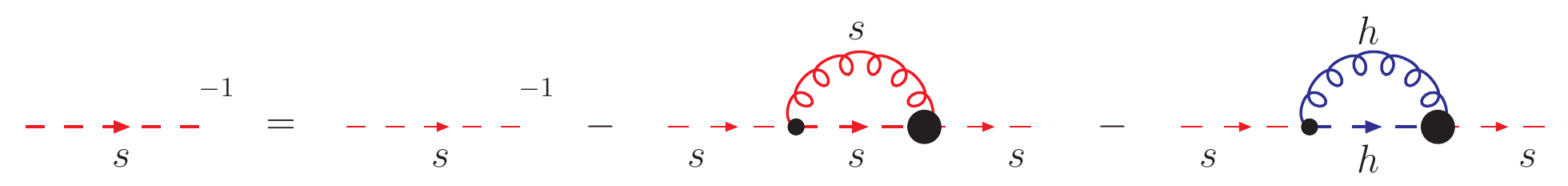}\end{center}

Since it depends only on a single momentum the dominant loop momenta
that contribute in the loop graph are normally of the order of the
soft external momentum scale $k\sim p$ but when masses are generated
dynamically also much larger momenta $k\sim\Lambda_{QCD}\gg p$ can
contribute and could in principle be relevant due to the positive
mass dimension of the graph. In particular, because of the possibility
of kinematic singularities of the vertices one obtains a nontrivial
contribution from large loop momenta that involve the IR exponent
of the ghost-gluon vertex in the limit that the ghost leg becomes
soft. Although the integration measure for this contribution from
hard momenta does not scale with the soft external momentum, it is
suppressed due to cancelations owing to the gluon transversality eq.
(\ref{eq:transversality}) which introduce a canonical scaling part
for the ghost-gluon vertices. This immediately rules out the possibility
of an IR constant ghost propagator since the leading contributions
from finite modes that could produce such an IR ghost mass $m_{gh}\left(p\right)\gg p$
are canceled in the ghost integral and only subleading contributions
remain. Because of the absence of such a mass $\mu_{gh}\equiv1$ and
there is the possibility of an additional suppression of the propagator
eq. (\ref{eq:hard-suppression}) given by $\mu_{gl}$ in case the
gluon propagator is likewise massless. The corresponding equation
for the anomalous dimension of the ghost reads therefore

\begin{equation}
-\delta_{gh}+1=\min\left(1;\delta_{gg}^{u}+\delta_{gh}+\delta_{gl}+1,\delta_{gg}^{gh}+1+\mu_{gl}\right)\:,\label{eq:soft-ghost-counting}\end{equation}
where here and in the following we separate terms arising from different
graphs by semicolons and those arising from different kinematic regions
of the same graph by commas. \\
The gluon equation truncated to IR leading 1-loop terms is given
by

\begin{center}\includegraphics[scale=0.7]{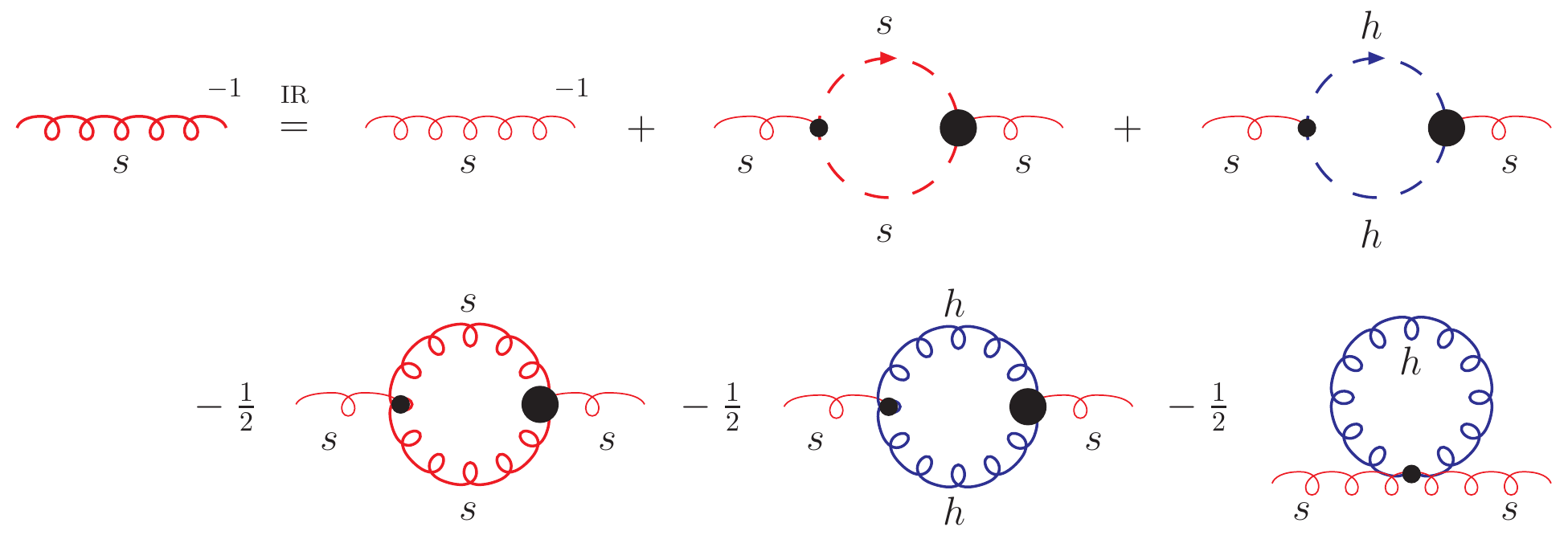}\end{center}

As in the ghost case there can be contributions from the hard part
of the loop integral. In particular, in this case there is no suppression
due to transversality as in the loop correction of the ghost propagator
and therefore an IR gluon mass is not ruled out. The absence of an
IR ghost mass implies that according to eq. (\ref{eq:hard-suppression})
the contribution from the hard region of the ghost loop is suppressed
by $p^{2}/k^{2}$ so that this correction cannot induce an IR gluon
mass. This is different for the gluonic corrections that can contribute
once a gluon mass has been dynamically generated but are likewise
suppressed otherwise. Since the external scale does not enter the
tadpole correction at all, it is not necessary to distinguish between
different kinematic regions in this case and as long as the there
is no scale induced this contribution is canceled identically. Correspondingly,
the gluon equation reads

\begin{equation}
-\delta_{gl}+1=\min\left(1;\delta_{gg}^{u}+2\delta_{gh}+1,\delta_{gg}^{gl}+1;\delta_{3g}^{u}+2\delta_{gl}+1,\delta_{3g}^{gl}+\mu_{gl};\mu_{gl}\right)\:,\end{equation}
The leading contributions in the ghost-gluon DSE are given by the
two triangle graphs that are analogous to the Abelian and non-Abelian
diagrams in the quark-gluon vertex. As discussed in section \ref{sec:IR-analysis},
when scales of different order of magnitude are involved, the decomposition
of the loop integrals yields in addition to a hard contribution generally
two contributions from inequivalent kinematic regions of the loop
integral shown in fig. \ref{fig:soft-loops} where different internal
momenta become soft. These correspond to inequivalent ways to route
the large momentum through the loop.\\

\begin{figure}
\begin{minipage}[c][1\totalheight][t]{0.27\textwidth}%
$\;$%
\end{minipage}%
\begin{minipage}[t]{0.45\textwidth}%
\includegraphics[scale=0.43]{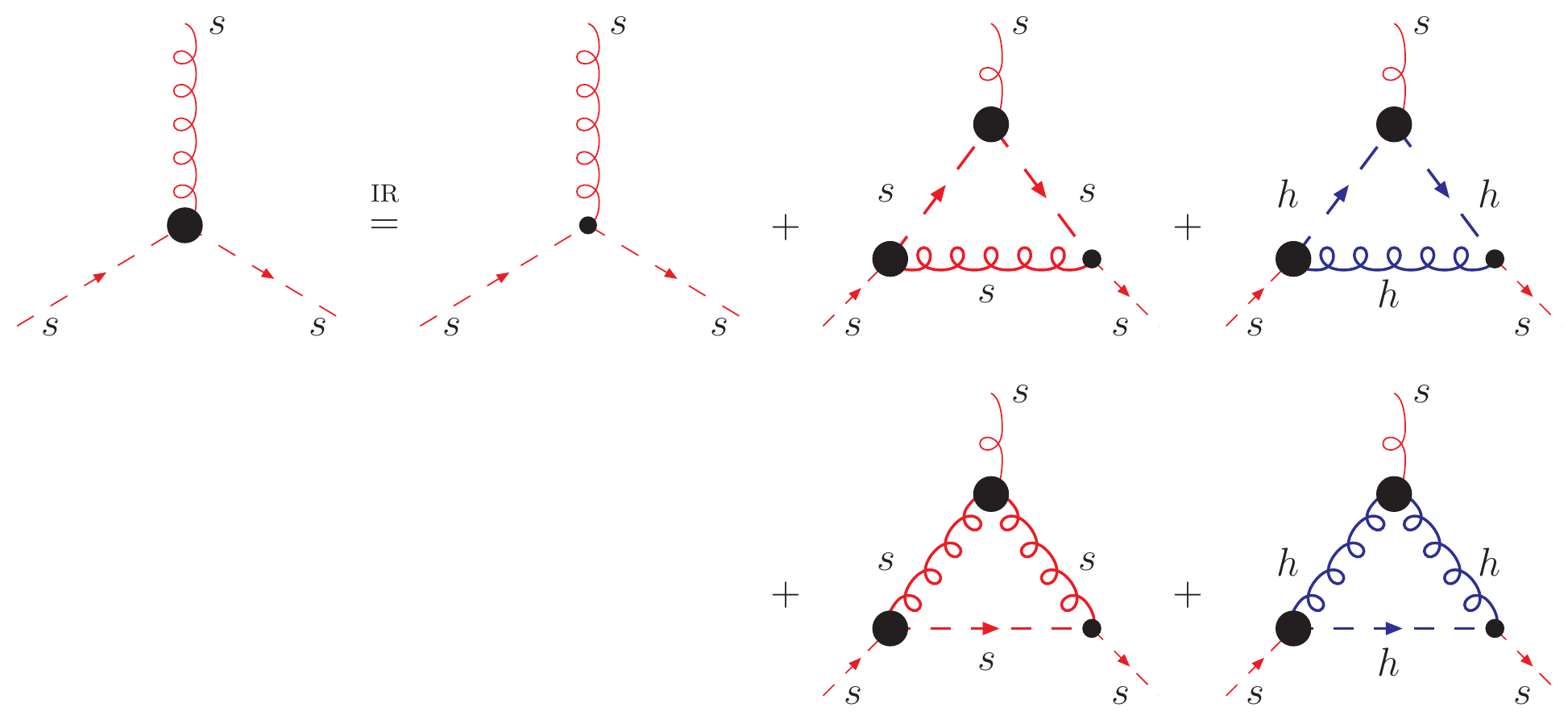}%
\end{minipage}%
\begin{minipage}[c][1\totalheight][t]{0.27\textwidth}%
$\;$%
\end{minipage}\\
\begin{minipage}[t]{0.48\textwidth}%
\includegraphics[scale=0.43]{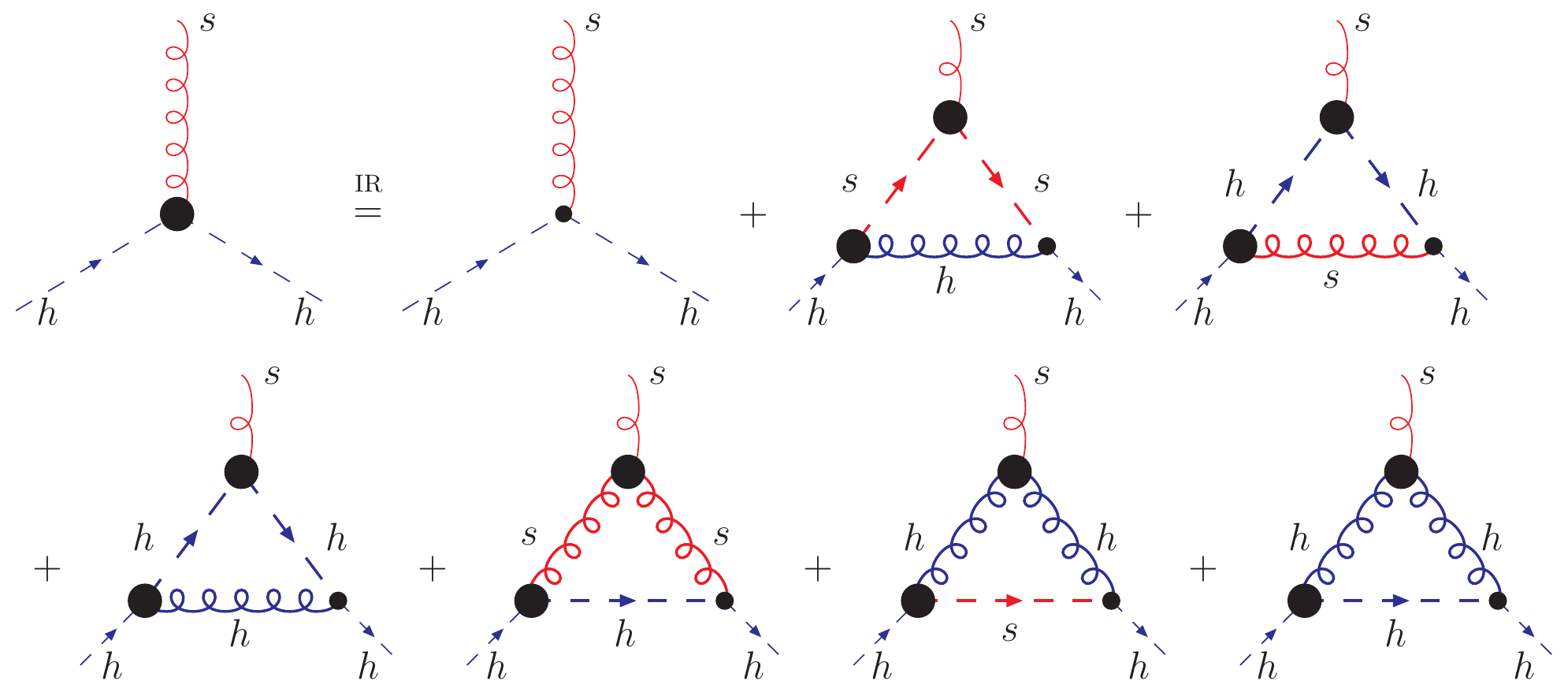}%
\end{minipage}%
\begin{minipage}[c][1\totalheight][t]{0.05\textwidth}%
$\;$%
\end{minipage}%
\begin{minipage}[t]{0.47\textwidth}%
\includegraphics[scale=0.43]{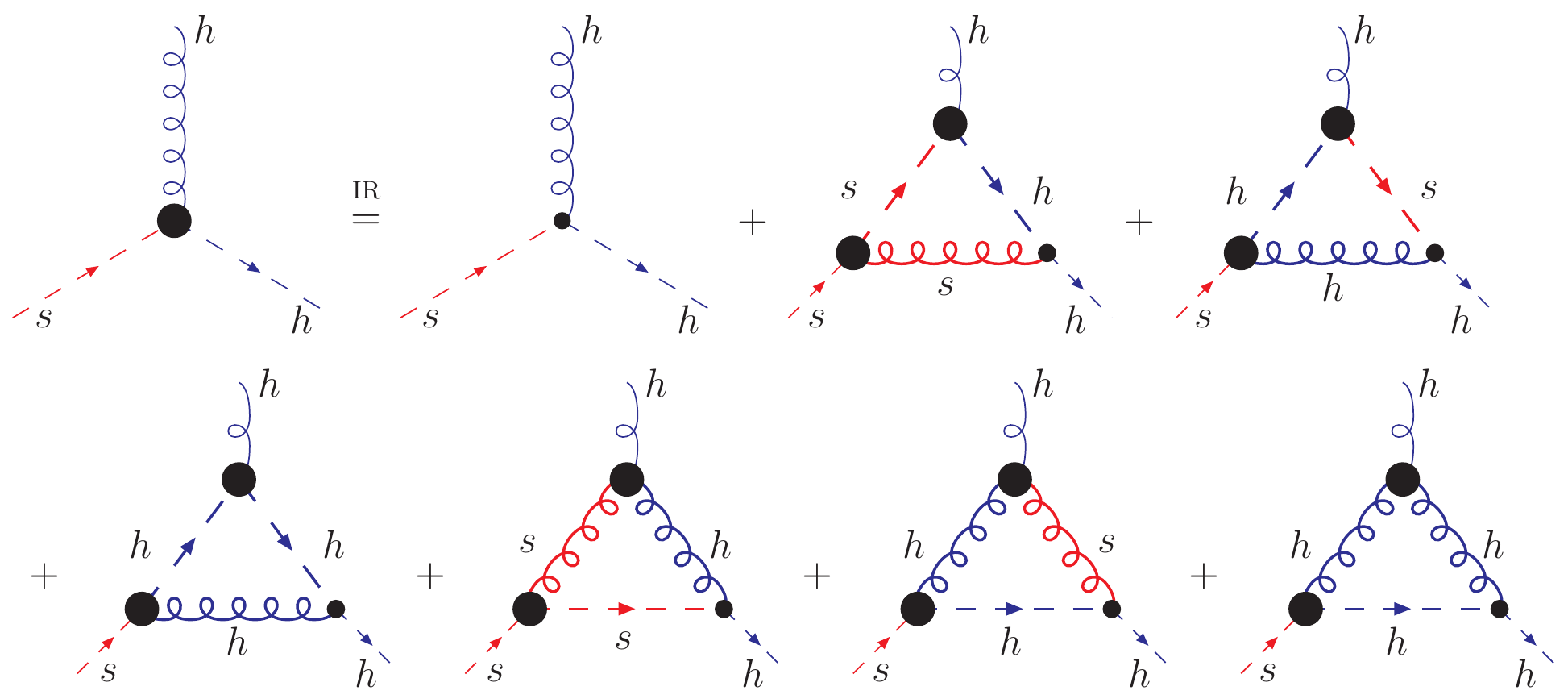}%
\end{minipage}

\caption{The IR leading part of the decomposed equations for the ghost-gluon
vertex in the uniform (top), soft-gluon (bottom left) and soft-ghost
limit (bottom right).\label{fig:ghost-gluon-kinematics}}

\end{figure}
E.g. for hard and soft external momenta $p$ and $q$, respectively,
and an assignment of the loop momentum $k$ such that the first IR
relevant contribution in fig. \ref{fig:soft-loops} arises from soft
loop momenta $k\sim q$, the second one arises from hard loop momenta
in a correspondingly narrow momentum interval $k-p\sim q\ll p\sim k$.
When assessing the counting of the vertex, each of these kinematic
contributions of the initial loop integral could dominate and has
to be taken into account separately. Using the graphical representation
depicted in figs. \ref{fig:soft-loops} and \ref{fig:uni-loops} the
ghost-gluon DSE in the different kinematic limits is explicitly represented
in fig. \ref{fig:ghost-gluon-kinematics}. We should stress that there
is no double counting involved here, since the figure only visualizes
the different distinct IR-sensitive regions of the loop integral that
arose from the exact decomposition discussed in section \ref{sec:IR-analysis}
and appendix \ref{sec:Decomposition}. \\
From the decomposed DSE we obtain the IR counting of the individual
contributions to the ghost-gluon vertex in the uniform limit

\begin{equation}
\delta_{gg}^{u}+\frac{1}{2}=\min\left(\frac{1}{2};2\delta_{gg}^{u}\!+\!2\delta_{gh}\!+\!\delta_{gl}\!+\!\frac{1}{2},\delta_{gg}^{gl}\!+\!\delta_{gg}^{gh}\!+\!1\!+\!\mu_{gl};\delta_{3g}^{u}\!+\!\delta_{gg}^{u}\!+\!\delta_{gh}\!+\!2\delta_{gl}\!+\!\frac{1}{2},\delta_{3g}^{gl}\!+\!\delta_{gg}^{gh}\!+\!1\!+\!\mu_{gl}\right)\:.\label{eq:uni-gg-counting}\end{equation}
Here in the minimum function the first element is the bare vertex,
the next two arise from the regions of soft and hard loop momenta
of the {}``Abelian'' graph whereas the final two are the contributions
from the corresponding regions of the {}``non-Abelian'' graph. In
the contributions from hard momenta the $1$ arises again from cancelations
due to the gluon transversality. In the limit that only the gluon
becomes soft we find

\begin{equation}
\delta_{gg}^{gl}=\min\left(0;\delta_{gg}^{u}\!+\!\delta_{gg}^{gh}\!+\!2\delta_{gh}\!+\!\frac{3}{2},2\delta_{gg}^{gl}\!+\!\delta_{gl}\!+\!1,\delta_{gg}^{gl};\delta_{3g}^{u}\!+\!\delta_{gg}^{gl}\!+\!2\delta_{gl}\!+\!\frac{1}{2},\delta_{3g}^{gl}\!+\!\delta_{gg}^{gh}\!+\!\delta_{gh}\!+\!2,\delta_{3g}^{gl}\!+\!\frac{1}{2}\right)\:.\label{eq:soft-gluon-gg-counting}\end{equation}
The additional suppression of the last term from the hard momentum
region of the {}``non-Abelian'' graph arises from the combination
of gluon transversality and the Bose symmetry of the 3-gluon vertex.
Choosing as the basis for the dressed ghost-gluon vertex the outgoing
ghost- and the gluon-momentum, $\Gamma_{gg}\!=\! A\, p_{gh}^{out}\!+\! B\, p_{gl}$,
with corresponding dressing functions $A$ and $B$, the contraction
of the dressed and bare ghost-gluon vertex with the corresponding
gluon propagators in the loop gives, respectively,

\begin{align}
\left(A\cdot\left(k+h+s\right)_{\gamma}+B\cdot\left(k+s\right)_{\gamma}\right)\left(\delta_{\gamma\alpha}-\frac{\left(k+s\right)_{\gamma}\left(k+s\right)_{\alpha}}{\left(k+s\right)^{2}}\right) & \approx A\cdot\left(h_{\alpha}-\frac{k\cdot h}{k^{2}}\, k_{\alpha}\right)+O\left(s\right)\:,\nonumber \\
\left(h+s\right)_{\delta}\left(\delta_{\delta\beta}-\frac{k_{\delta}k_{\beta}}{k^{2}}\right) & \approx h_{\beta}-\frac{k\cdot h}{k^{2}}\, k_{\beta}+O\left(s\right)\:,\end{align}
where $h$ is the incoming hard external ghost momentum, $s$ the
soft external gluon momentum and $k$ the loop momentum. Because of
the antisymmetry of the color part, the Bose symmetry of the dressed
3-gluon vertex implies that it is antisymmetric with respect to simultaneous
commutation of momenta and Lorentz indices. The appearance of two
identical momenta up to $O\left(s\right)$ corrections shows then
that the leading term vanishes and the loop scales actually as $O\left(s\right)$.
The corresponding equation in the soft ghost limit follows similarly
from the power counting analysis

\begin{equation}
\delta_{gg}^{gh}=\min\left(0;\delta_{gg}^{u}\!+\!\delta_{gg}^{gh}\!+\!\delta_{gh}\!+\!\delta_{gl}\!+\!\frac{1}{2},2\delta_{gg}^{gh}\!+\!\delta_{gh}\!+\!2,\delta_{gg}^{gh};\delta_{gg}^{u}\!+\!\delta_{3g}^{gl}\!+\!\delta_{gh}\!+\!\delta_{gl}\!+\!1,\delta_{3g}^{gl}\!+\!\delta_{gg}^{gh}\!+\!\delta_{gl}\!+\!\frac{3}{2},\delta_{gg}^{gh}\right)\:.\label{eq:soft-ghost-gg-counting}\end{equation}
\begin{figure}
\begin{minipage}[t]{0.48\textwidth}%
\includegraphics[scale=0.43]{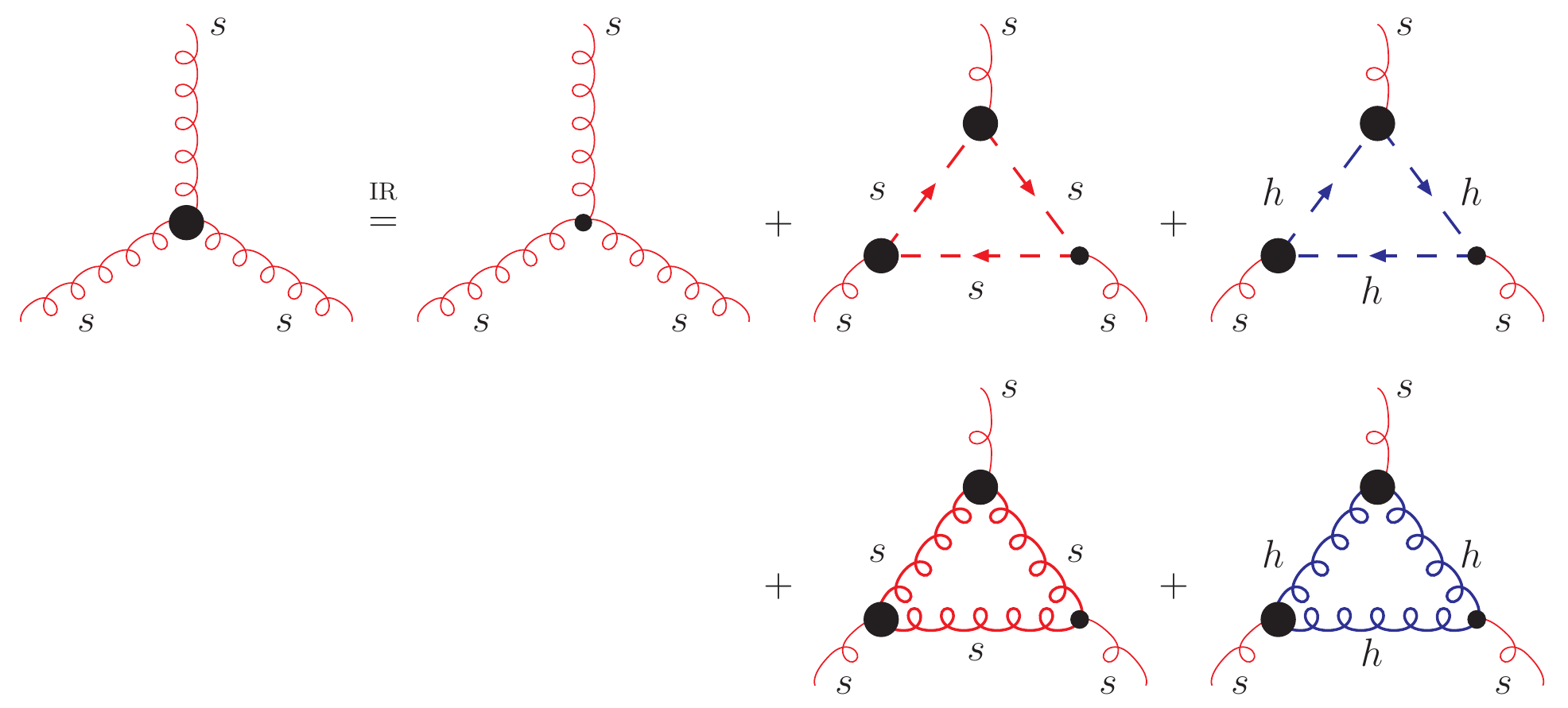}%
\end{minipage}%
\begin{minipage}[c][1\totalheight][t]{0.05\textwidth}%
$\;$%
\end{minipage}%
\begin{minipage}[t]{0.47\textwidth}%
\includegraphics[scale=0.43]{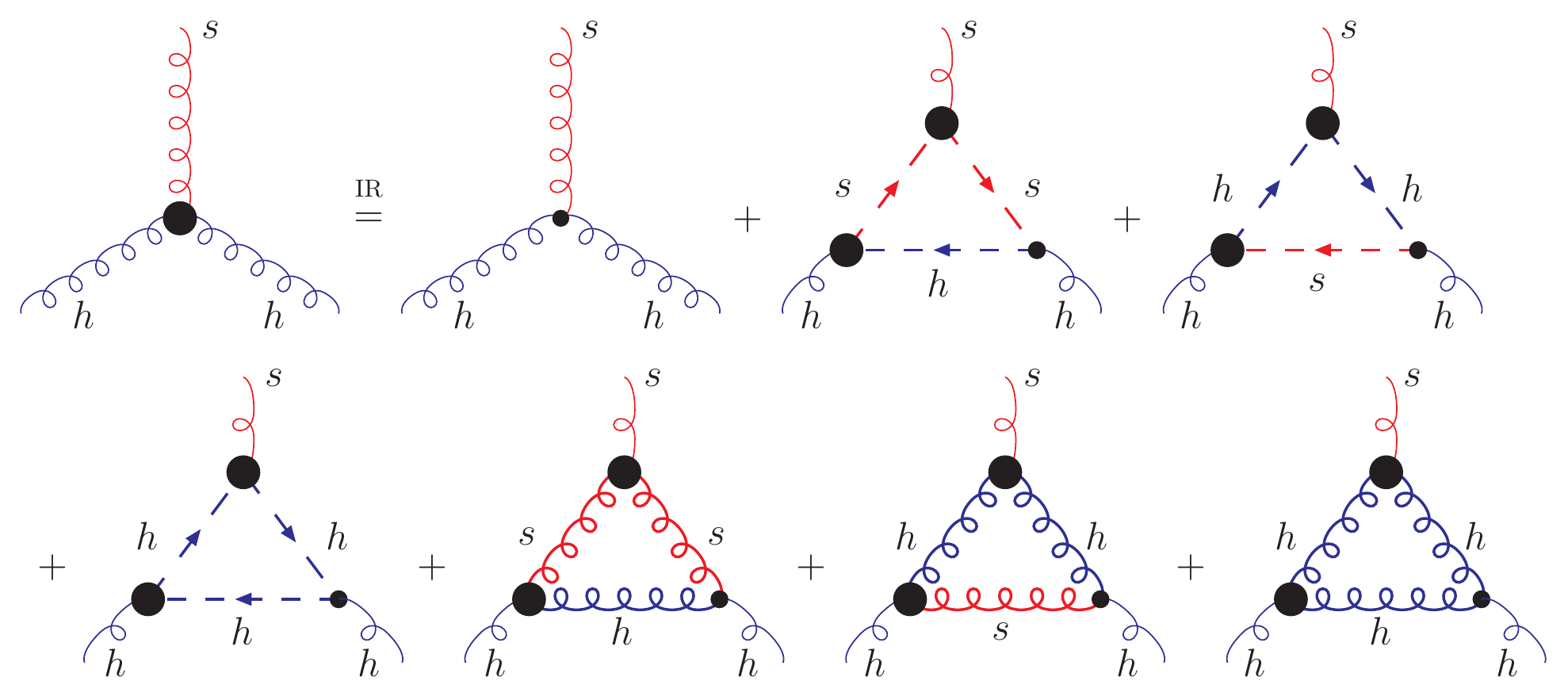}%
\end{minipage}

\caption{The IR leading part of the decomposed equations for the 3-gluon vertex
in the uniform (left) and soft-gluon limit (right).\label{fig:3-gluon-kinematics}}

\end{figure}

Finally in the considered truncation there is the equation for the
3-gluon vertex where the contributions are given by the ghost and
gluon triangles. The equations for the two different kinematic limits
are shown in fig. \ref{fig:3-gluon-kinematics}. In the uniform limit
the power counting yields the equation

\begin{equation}
\delta_{3g}^{u}+\frac{1}{2}=\min\left(\frac{1}{2};2\delta_{gg}^{u}+3\delta_{gh}+\frac{1}{2},2\delta_{gg}^{gl}+1;2\delta_{3g}^{u}+3\delta_{gl}+\frac{1}{2},2\delta_{3g}^{gl}+\frac{1}{2}+\mu_{gl}\right)\:,\end{equation}
where the $1/2$ in the contribution from hard modes of the gluon
triangle arises since the leading term depending only on the loop
momentum is odd and vanishes in the symmetric integration. The soft-gluon
limit requires some more care. Here, the IR exponent of the ghost
triangle seems to depend on different factors like the loop routing,
the definition of the bare ghost vertex or which of the three vertices
is taken bare in the DSE. All these factors seemingly determine whether
the appearing bare ghost-gluon vertex scales canonically. To see that
this is actually the case independent of all these conventions, it
is important to remember that the 3-gluon vertex is totally symmetric.
This means that all three ways of assigning the external momenta and
indices to the external legs of the graph yield the same result. With
the standard convention that the bare vertex is proportional to the
outgoing ghost momentum, this momentum is soft for at least one of
the three configurations. Since the IR exponent obtained by the power
counting analysis can overestimate the degree of divergence of a
given graph when there are cancelations but it cannot underestimate
it, the bare ghost-gluon vertex features indeed a canonical scaling.
Correspondingly one obtains here

\begin{equation}
\delta_{3g}^{gl}=\min\left(0;\delta_{gg}^{u}\!+\!\delta_{gg}^{gh}\!+\!2\delta_{gh}\!+\!1,\delta_{gg}^{gl}\!+\!\delta_{gg}^{gh}\!+\!\delta_{gh}\!+\!\frac{3}{2},\delta_{gg}^{gl};\delta_{3g}^{u}\!+\!\delta_{3g}^{gl}\!+\!2\delta_{gl}\!+\!\frac{1}{2},2\delta_{3g}^{gl}\!+\!\delta_{gl}\!+\!1,\delta_{3g}^{gl}\right)\:.\label{eq:soft-3g-counting}\end{equation}
The fourth element in the minimum function gives a constraint on
the two vertices in the soft gluon limit $\delta_{3g}^{gl}\leq\delta_{gg}^{gl}$.
It is straightforward to see that the reduced system of the three
equations for the soft divergences alone is consistent with the uniform
solution discussed in the last section if the corresponding IR exponents
are taken as given. Though, here we want to discuss the general case
where the additional kinematic divergences couple back and could thereby
change the uniform solution. To this end we study the full, coupled
system of seven equations obtained by the consideration of possible
kinematic divergences.\\
We want to stress that, analogous to the case discussed in detail
in section \ref{sec:IR-slavery}, the IR exponents on the left-hand
side in the vertex eqs. (\ref{eq:uni-gg-counting}), (\ref{eq:soft-gluon-gg-counting}),
(\ref{eq:soft-ghost-gg-counting}) and (\ref{eq:soft-3g-counting})
appear linearly also on the right-hand side. The corresponding inequalities,
e.g. $\delta_{gg}^{u}+1/2\leq\delta_{gg}^{u}+\delta_{3g}^{u}+\delta_{gh}+2\delta_{gl}+1/2$
from eq. (29), can be fulfilled independently of the values of the
IR exponents on the left-hand side, since they drop out. Yet, they
provide constraints on the other IR exponents. From a purely mathematical
point of view these DSEs could be trivially fulfilled for a whole
range of values for the corresponding IR exponent if these linear
terms would dominate. Yet, such linear terms provide no non-linear
feedback and cannot alter the degree of divergence compared to the
tree level term in the DSE that is not IR enhanced. Correspondingly
from a physical point of view these terms are irrelevant for the IR
solution.\\
As in the uniform case the system of IR equations is again constrained
by necessary conditions for a stable skeleton expansion. In addition
to those in the uniform limit eqs. (\ref{eq:se-rel}) there are corresponding
constraints from the graphs in fig. \ref{fig:sekelton} when the loop
momentum is in the IR regime but one of the two connected propagators
has a hard momentum 

\begin{align}
\delta_{gg}^{u}+\delta_{gg}^{gl}+\delta_{gh}+\delta_{gl}+\frac{1}{2} & \geq0\:,\nonumber \\
\delta_{3g}^{u}+\delta_{gg}^{gl}+2\delta_{gl}+\frac{1}{2} & \geq0\:,\nonumber \\
\delta_{gg}^{u}+\delta_{3g}^{gl}+\delta_{gh}+\delta_{gl}+\frac{1}{2} & \geq0\:,\nonumber \\
\delta_{3g}^{u}+\delta_{3g}^{gl}+2\delta_{gl}+\frac{1}{2} & \geq0\:,\label{eq:soft-constraints}\end{align}
 or when the momenta of both connected propagators are hard

\begin{align}
2\delta_{gg}^{gl}+\delta_{gl}+1 & \geq0\:,\nonumber \\
\delta_{3g}^{gl}+\delta_{gg}^{gl}+\delta_{gl}+1 & \geq0\:,\nonumber \\
2\delta_{3g}^{gl}+\delta_{gl}+1 & \geq0\:.\label{eq:hard-constraints}\end{align}
In principle there may also be constraints from the hard part of the
added loop, but in this case the extension can change the counting
far away from the insertion and therefore they do not take a simple
form. Using the constraints eqs. (\ref{eq:se-rel}) , (\ref{eq:soft-constraints})
and (\ref{eq:hard-constraints}) as well as $\delta_{3g}^{gl}\leq\delta_{gg}^{gl}$,
the system of inequalities, eqs. (\ref{eq:soft-ghost-counting})-(\ref{eq:soft-3g-counting}),
reduces to:

\begin{align}
-\delta_{gh}+1 & =\min\left(1,\delta_{gg}^{u}+\delta_{gh}+\delta_{gl}+1,\delta_{gg}^{gh}+1+\mu_{gl}\right)\:,\nonumber \\
-\delta_{gl}+1 & =\min\left(1,\delta_{gg}^{u}+2\delta_{gh}+1,\delta_{3g}^{u}+2\delta_{gl}+1,\delta_{3g}^{gl}+\mu_{gl}\right)\:,\nonumber \\
\delta_{gg}^{u}+\frac{1}{2} & =\min\left(\frac{1}{2},\delta_{3g}^{gl}\!+\!\delta_{gg}^{gh}\!+\!1\!+\!\mu_{gl}\right)\:,\nonumber \\
\delta_{3g}^{u}+\frac{1}{2} & =\min\left(\frac{1}{2},2\delta_{gg}^{u}+3\delta_{gh}+\frac{1}{2},2\delta_{3g}^{gl}+\frac{1}{2}+\mu_{gl}\right)\:,\nonumber \\
\delta_{gg}^{gh} & =\min\left(0,2\delta_{gg}^{gh}\!+\!\delta_{gh}\!+\!2\right)\:,\nonumber \\
\delta_{gg}^{gl} & =\min\left(0,\delta_{gg}^{u}\!+\!\delta_{gg}^{gh}\!+\!2\delta_{gh}\!+\!\frac{3}{2},\delta_{3g}^{gl}\!+\!\delta_{gg}^{gh}\!+\!\delta_{gh}\!+\!2,\delta_{3g}^{gl}\,+\,\frac{1}{2}\right)\:,\nonumber \\
\delta_{3g}^{gl} & =\min\left(0,\delta_{gg}^{u}\!+\!\delta_{gg}^{gh}\!+\!2\delta_{gh}\!+\!1,\delta_{gg}^{gl}\!+\!\delta_{gg}^{gh}\!+\!\delta_{gh}\!+\!\frac{3}{2},\delta_{gg}^{gl}\right)\:.\label{eq:soft-system}\end{align}
In the uniform case the non-renormalization of the ghost-gluon vertex
was the cornerstone that allowed one to solve the corresponding system.
We therefore start with this vertex in the soft ghost limit described
by the equation for $\delta_{gg}^{gh}$. Let us assume for the moment
that the vertex is singular so that the second term dominates. In
this case we would have $\delta_{gg}^{gh}=-\delta_{gh}-2$, but this
would yield a direct contradiction via the last term of the ghost
propagator equation. Therefore we have instead $\delta_{gg}^{gh}=0$
and obtain additionally the weak constraint $\delta_{gh}\geq-2$.
In contrast to the conformal case the remaining system allows two
qualitatively different solutions:

\paragraph*{Decoupling solution:}

With any renormalization prescription that does not change the above
system for the IR exponents, as in the conformal case discussed above
the equation for the ghost propagator as it stands has only the trivial
solution $\delta_{gh}=0$. To find the solution of the residual system
in the present case, we insert the remaining possible solutions for
the 3-gluon vertex in the soft gluon limit into the other equations
which yields

\begin{equation}
-\delta_{gl}+1=\min\left(1,\mu_{gl}\right)\:,\quad\delta_{gg}^{u}=\delta_{gg}^{gl}=\delta_{3g}^{u}=\delta_{3g}^{gl}=0\:.\end{equation}
In addition to the trivial fixed point, there is here also the possibility
of a massive gluon $\delta_{gl}=1$. Therefore, in contrast to the
conformal analysis where it was assumed that all integrals are dominated
by scales of the order of the external momenta, hard modes of the
order of the induced scale $\Lambda_{QCD}$ can dominate the loop
integrals in the equation for the gluon propagator and induce an IR
gluon mass. This alternative scenario is reminiscent of the early
work \cite{Cornwall:1981zr} and has recently also been suggested
in \cite{Boucaud:2008ji,Aguilar:2008xm,Fischer:2008uz} on the level
of the propagator equations, whereas the present analysis shows that
it is also consistent with the vertex equations. This solution is
also found if the ghost propagator is renormalized to a finite value
at vanishing momentum \cite{Fischer:2008uz}.

\paragraph*{Scaling solution:}

Alternatively, it is again possible to use the renormalization introduced
in \cite{vonSmekal:1997is,Lerche:2002ep} as discussed in the last
section which removes the tree-level term in the ghost equation. Because
of the scale independence of the ghost-gluon vertex in the limit that
a ghost momentum vanishes the last term in the ghost equation is either
directly subleading in case the gluon propagator remains massless
or can otherwise be removed in the renormalization process together
with the tree-level term. Thereby the ghost equation becomes unique

\begin{equation}
\delta_{gh}=-\frac{1}{2}\left(\delta_{gl}+\delta_{gg}^{u}\right)\end{equation}
and inserting this expression for the ghost exponent, the gluon equation
becomes trivially fulfilled

\begin{equation}
-\delta_{gl}+1=\min\left(1,-\delta_{gl}+1,\delta_{3g}^{u}+2\delta_{gl}+1,\delta_{3g}^{gl}+\mu_{gl}\right)\,,\end{equation}
which shows that the ghost-loop has to be the IR leading contribution.
It does not provide any constraint on $\delta_{gl}$, leaving the
free parameter $\kappa\equiv\delta_{gl}/2\geq0$ in the solution as
is known already from the uniform case. We note that the triviality
of this equation is qualitatively different from what we found in
the case of the equations for the kinematic divergences above, since
the equation is non-linear and can therefore indeed self-consistently
generate a non-trivial solution. The equation provides two additional
weak constraints on the vertices $\delta_{3g}^{u}\geq-6\kappa$, $\delta_{3g}^{gl}\geq1\!-\!2\kappa\!-\!\mu_{gl}$.
Inserting the above expression for the ghost exponent in the first
of the constraints from the skeleton expansion in the uniform limit
eqs. (\ref{eq:se-rel}) yields $\delta_{gg}^{u}\geq0$ which makes
the corresponding equation for the exponent of the uniform ghost-gluon
vertex unique, so that $\delta_{gg}^{u}=0$. The residual system reads
now

\begin{align}
\delta_{3g}^{u}+\frac{1}{2} & =\min\left(-3\kappa+\frac{1}{2},2\delta_{3g}^{gl}+\frac{1}{2}+\mu_{gl}\right)\:,\nonumber \\
\delta_{gg}^{gl} & =\min\left(0,-2\kappa\!+\!\frac{3}{2},\delta_{3g}^{gl}\!-\!\kappa\!+\!2,\delta_{3g}^{gl}\,+\,\frac{1}{2}\right)\:,\nonumber \\
\delta_{3g}^{gl} & =\min\left(0,-2\kappa\!+\!1,\delta_{gg}^{gl}\!-\!\kappa\!+\!\frac{3}{2},\delta_{gg}^{gl}\right)\:.\end{align}
Using the constraint $\delta_{3g}^{gl}\leq1\!-\!2\kappa$ from the
last of these equations in the third of eqs. (\ref{eq:soft-constraints})
yields the constraint $\kappa<3/2$, so that a solution can only exist
in a bounded region of the IR scaling parameter $\kappa$. The system
can then be solved by inserting the possible solutions of the ghost-gluon
equation in the soft-gluon limit in the equation for the 3-gluon vertex
in the same limit. Under consideration of this bound the equation
reduces to $\delta_{3g}^{gl}=\min\left(0,-2\kappa\!+\!1\right)$.
Correspondingly, the scaling of the $3$-gluon vertex in the soft-gluon
limit depends on the value of the IR parameter $\kappa$. For $\kappa>1/2$
it is singular $\delta_{3g}^{gl}=1-2\kappa$ whereas for $\kappa\leq1/2$
it is not $\delta_{3g}^{gl}=0$. Using these results the remaining
equation becomes trivial and the \emph{scaling solution} of the system
is obtained. Similar to the 3-gluon vertex the exponent in the soft-gluon
limit of the ghost-gluon vertex $\min\left(0,\frac{3}{2}-2\kappa\right)$
also depends on $\kappa$, but here a divergence would arise only
for $\kappa\geq3/4$. Finally, the scaling of the 4-gluon vertex
in the uniform limit that satisfies the DSE system is obtained from
the corresponding DSE fig. \ref{fig:4-gluon-DSE} as before and yields
the known result $\delta_{4g}^{u}=-4\kappa$ whereas the determination
of the corresponding kinematic divergences requires a more detailed
study. In summary, the IR fixed points for the two qualitative distinct
solutions are given by Table \ref{tab:IR-scaling-in}.

\begin{table}[h]
\begin{tabular}{|c|c|c|c|c|c|c|c|c|}
\hline 
 & $\delta_{gh}$ & $\delta_{gl}$ & $\delta_{gg}^{u}$ & $\delta_{3g}^{u}$ & $\delta_{4g}^{u}$ & $\delta_{gg}^{gh}$ & $\delta_{gg}^{gl}$ & $\delta_{3g}^{gl}$\tabularnewline
\hline 
scaling & $-\kappa$ & $2\kappa$ & $0$ & $-3\kappa$ & $-4\kappa$ & $0$ & $\min\left(0,\tfrac{3}{2}-2\kappa\right)$ & $\min\left(0,1-2\kappa\right)$\tabularnewline
\hline
decoupling & $0$ & $1$ & $0$ & $0$ & $0$ & $0$ & $0$ & $0$\tabularnewline
\hline
\end{tabular}

\caption{\label{tab:IR-scaling-in}The IR exponents for the leading Green functions
of the IR fixed points of Landau gauge Yang-Mills theory within the
two possible IR scenarios. The scaling analysis restricts the parameter
to be only positive $\kappa\geq0$, and yields rather weak upper bounds.
(Note that these bounds on $\kappa$ are weaker than the one that
had been erroneously given in a preprint version of this article.
Stronger bounds $0.5\leq\kappa\leq1$ are, however, supported by analyses
of the actual loop integrals \cite{Zwanziger:2001kw}.)}

\end{table}

These solutions fulfill all constraints that appeared in the course
of the evaluation and present therefore refined IR fixed points of
Landau-gauge Yang-Mills theory. Several remarks are in order at this
point:
\begin{itemize}
\item In this work we restricted our analysis to the IR behavior of Yang-Mills
theory. The physical relevance of such a study stems certainly from
the fact that Yang-Mills theory presents the gauge sector of QCD and
might therefore provide important insight into qualitative properties
of the strong interaction. Whereas an independent study of the gauge
dynamics is by definition sufficient in the quenched limit \cite{Alkofer:2008tt},
in the case of dynamical QCD it is not \emph{a priori} clear that
the quark dynamics does not affect the IR fixed point structure of
gauge Green functions obtained here. Yet, a corresponding recent IR
analysis of QCD \cite{Schwenzer:2008vt}, based on the general methods
developed in this work, shows that the gauge sector is totally unaltered
by the quark dynamics and thereby strongly supports the relevance
of the results presented here and in other studies of Yang-Mills theory.
\item In contrast to the case of the conformal analysis discussed in the
previous section, where the chosen boundary condition merely excluded
other solutions, the decoupling solution obtained for a generic condition
is qualitatively different from the scaling solution obtained for
a choice that had an unbroken global BRS charge \cite{Fischer:2008uz}.
Whereas in the scaling solution the ghosts are strongly IR enhanced
resulting in divergent gluonic vertices, in the decoupling solution
neither the ghosts nor the vertices are anomalously enhanced. This
strongly suppresses any IR dynamics mediated by the gluons in the
ratio $\rho\equiv p^{2}/m_{gl}^{2}$ where $m_{gl}$ is the finite
IR limit of the gluon polarization. Thereby it is easy to see that
in both cases the leading contribution to a Green function is given
by the ghost dynamics. In the decoupling case the IR exponents in
Table \ref{tab:IR-scaling-in} show directly that the leading term
in the skeleton expansion of a general vertex with $n$ ghost-pairs
is anomalously suppressed by $\rho^{n}$ whereas purely gluonic Green
functions scale canonically. In addition to the given decoupling solution
there might be further IR fixed points where even the vertices decouple
and become IR constant \cite{Jan}. 
\item Recent lattice simulations on large lattices \cite{Bowman:2007du}
show a gluon propagator that does not feature a decrease in the IR
and a ghost propagator that is basically not IR enhanced and thereby
favor the decoupling scenario. Moreover, it has been argued that this
is probably neither a finite volume \cite{Fischer:2007pf} nor a
statistical effect \cite{Cucchieri:2007rg}. However, there seem
to be issues with Gribov copies \cite{Sternbeck:2005tk,Maas:2008ri}
and discretization effects \cite{Sternbeck:2008mv} in these analyses.
Such effects could shadow a potential scaling behavior in the deep
IR, so that the scaling scenario cannot be excluded with the present
data. Another view is presented in \cite{Maas:2009se}, where it
is suggested that the different boundary conditions \cite{Fischer:2008uz}
correspond to distinct residual gauge fixing conditions. In principle
Gribov copies could be an issue in the DSE system as well and the
constraint to the fundamental modular region that is free of Gribov
copies could change these equations and their solution structure.
However, it has recently been shown explicitly \cite{Huber:2009tx}
that at least the restriction to the first Gribov region \cite{Gribov:1977wm}
using the Gribov-Zwanziger action \cite{Zwanziger:1989mf} does not
affect the IR fixed point structure of the scaling solution.
\item The IR exponents given in Table \ref{tab:IR-scaling-in} determine
only the anomalous scaling laws for the most singular tensor parts.
The scaling of the full Green functions involves also the canonical
scaling dimension incorporated in the tensors. In particular it
is possible that some dressing functions are more IR singular than
the leading dressing function, but their contribution to the vertex
is nevertheless subleading since it is additionally suppressed by
the canonical scaling of their tensor. As we show in \cite{Alkofer:2008dt}
this is indeed the case for the ghost-gluon vertex which features
more structure than the above result suggests. Instead, a soft-gluon
singularity appears in the form factor of the longitudinal tensor
that is additionally suppressed by the gluon momentum in the tensor
and actually IR vanishing, whereas the tree-level tensor is entirely
IR finite and presents the IR leading structure. In order to reveal
such subtleties in our power counting analysis we would have had to
include different anomalous dimensions for the different tensor structures.
Since we present an explicit analytic solution for the IR limit of
the 3-point vertices in \cite{Alkofer:2008dt} we refrained here
from such complications. In \cite{Alkofer:2008dt} we show explicitly
that the transverse part of the 3-gluon vertex does not have leading
tensors. The analytic result obtained there supports this observation.
It also explains why the kinematic singularity for the three-point
vertices obtained in the second work of ref. \cite{Fischer:2006vf},
where only the transverse part was considered, is lower as the one
given in tab. \ref{tab:IR-scaling-in}.
\item The kinematic singularities do not alter but merely extend the previously
know uniform scaling fixed point. Yet, the soft singularities restrict
the range of possible $\kappa$-values from the mere positivity requirement
in the conformal case to the bounded interval $0\leq\kappa\leq3/2$.
 The trivial solution is included in the scaling solution for $\kappa=0$.
It is (up to logarithmic corrections) realized in the UV regime of
the theory characterized by asymptotic freedom and it is clear from
the perturbative $\beta$-functions that this solution can hardly
be a stable IR fixed point, too. The best currently known value for
the IR scaling parameter is $\kappa\approx0.5953$ \cite{Zwanziger:2001kw,Lerche:2002ep}
obtained from an analytic IR solution of the integrals in the DSEs
for the propagators.
\item Interestingly below $\kappa=1/2$ the kinematic singularities entirely
disappear. The latter value is a special case since the gluons show
effectively a massive behavior but the ghost is in contrast to the
corresponding decoupling solution still strongly divergent. For $\kappa>1/2$
the gluon propagator should vanish with the small exponent $2\kappa-1$
which is precisely the negative of the exponent for the mild kinematic
singularities of the 3-gluon vertex found here. In four dimensions
such a small exponent naturally poses a huge numerical challenge and
is not observed in current studies \cite{Bowman:2007du}, but in
lower dimensions a corresponding decrease has been clearly confirmed
\cite{Maas:2007uv}. Similarly, it is not surprising that the predicted
kinematic singularities have not been seen so far in present vertex
studies \cite{Cucchieri:2006tf} which are numerically even much
more challenging than those for the propagators. In \cite{Alkofer:2008dt}
we proposed a suitable tensor contraction for the three-gluon vertex
to investigate the existence of kinematic singularities on the lattice.
\item It is crucial that the ghost-gluon vertex is finite when only a ghost
momentum vanishes. This result follows immediately from the corresponding
{}``un-decomposed'' DSE which contains only a single loop graph
involving the connected (instead of 1PI) ghost-gluon scattering kernel
\cite{Alkofer:2004it}. By transversality this graph is directly
proportional to the external momentum and leaves only the tree level
part in the IR limit in accordance with the non-renormalization of
this vertex. We point out that there is, however, no corresponding
argument when the gluon momentum vanishes.
\item The obtained divergence when only a single gluon momentum vanishes
naively seems to be problematic for several reasons: First of all
it seems to induce an even stronger singularity in the ghost-gluon
vertex in the uniform limit from hard loop momenta. As pointed out
above though in this case the transversality in Landau gauge prevents
this and instead makes this contribution strongly subleading. This
is also in accordance with the two different versions for the ghost-gluon
vertex. Since a dressed 3-gluon vertex is present only in the first
one, the two versions would be inconsistent if the kinematic-divergence
of the 3-gluon vertex would alter the degree of divergence of the
full vertex. Secondly, naively there seems to be a huge problem with
the soft gluon singularity in the 3-gluon vertex. First of all it
arises directly from the ghost loop integral with dressed propagators.
But once induced, it seems to arise in addition also in dressed vertices
whenever the external momentum becomes soft and totally independent
of the loop integral. This would enhance the divergence in each iteration
and make it more and more divergent. As seen explicitly in the above
analysis the reason why this is not the case is that the hard region
is additionally suppressed and that both of these different singularities
arise from distinct regions of the loop integration and thereby cannot
amplify each other.
\item As found from the analysis above, all Green functions in the scaling
solution include a graph that does not involve singular vertices.
In particular, the IR dominant ghost loop correction to the 3-gluon
vertex induces the soft-gluon divergence entirely due to the enhancement
of the ghost propagator so that the appearance of kinematic divergences
is a direct consequence of the ghost dominance of the uniform solution.
This allows one to capture the qualitative IR behavior of the vertices
in a semi-perturbative scheme that involves dressed propagators but
employs bare vertices. This approximation is used in a companion article
and allows a complete analytic solution in terms of hypergeometric
functions \cite{Alkofer:2008dt}. Although the analysis of the kinematic
divergences of the 4-gluon vertex is more complicated and requires
a detailed analysis, we note here that the semi-perturbative contribution
from the ghost loop yields $\delta_{4g}^{gl}\!=\!1\!-\!2\kappa$ and
$\delta_{4g}^{2g}\!=\!1/2\!-\!3\kappa$ which due to the ghost dominance
is expected to be the leading contribution. Incidentally this is exactly
the scaling of the 3-gluon vertex in the soft-gluon and uniform limit
whereas the uniform limit of the 4-gluon vertex $\delta_{4g}^{u}\!=\!-4\kappa$
is more divergent by yet another $-1/2\!-\!\kappa$. These results
suggest the conjecture that the full scaling of a gluonic correlation
function with $n_{s}$ independent soft external momenta should be
$\Gamma_{n_{s}}\sim\left(p^{2}\right)^{2-\left(n_{s}+1\right)\left(1/2+\kappa\right)}$.
\item Inserting the results for the scaling solution in Table \ref{tab:IR-scaling-in}
in the constraints for the skeleton expansion involving hard momenta
we find that the extensions count as $3/2-\kappa$ and $3-2\kappa$
when there are one respectively two hard propagators in the extended
graph. With the above limits for $\kappa$ this shows that these extensions
are strongly suppressed and correspondingly such extensions do not
have to be taken into account in the skeleton expansion. In contrast,
inserting the results in the constraints for the skeleton expansion
in the uniform limit eqs. (\ref{eq:se-rel}), it is clear that all
of them are saturated and correspondingly all orders in the expansion
scale equally \cite{Alkofer:2004it}. Note at this point, that we
did not assume by our constraints eqs. (\ref{eq:se-rel}) that the
skeleton expansion strictly converges, but only that it is \emph{not
explicitly divergent}. In general, the skeleton expansion could be
an asymptotic series as suggested by the IR scaling. Therefore, the
whole tower of such graphs had to be resummed which could in principle
change the IR scaling. For instance it is well known from standard
resummed perturbation theory that the resummation of perturbative
logarithms yields a power law scaling with an anomalous exponent.
However, even such logarithmic divergences can be invariant under
resummation, as e.g. found for the non-Fermi liquid corrections in
dense QCD \cite{Schafer:2005mc}. In the current case a resummation
seems to be impossible in full generality, anyhow, but the decisive
difference is that the graphs that are resummed already feature power
law scaling and therefore we expect that they are indeed invariant
under resummation. Furthermore, it has been shown in \cite{Huber:2009wh,Alkofer:2006xz,Fischer:2006vf}
that the uniform IR solution can be obtained independently of the
skeleton expansion and the same will be shown for the more general
case as discussed in a forthcoming publication \cite{Skeleton,Schwenzer:2008vt}.
\item More generally the validity of the skeleton expansion is closely linked
to the existence of any finite truncation of the DSEs and eventually
to the concept of locality. To see this, note that an explicitly divergent
skeleton expansion suggests that there is no finite approximation
to describe higher order Green functions by lower ones or in particular
only in terms of the primitively divergent correlators. This would
mean that the higher Green functions include important physics that
is not yet included in the lower Green functions and that is required
to properly describe the system. In particular these higher Green
functions in turn significantly influence the solution of the leading
Green functions within an explicit analysis of the dynamics in this
case. Any truncation of the system with a local effective action that
includes only a finite number of terms would thereby miss the main
physics and instead a non-local effective action with an infinite
number of terms in the local fields is necessary. Such a non-local
situation means that the description in terms of local degrees of
freedom breaks down and the system can probably be more conveniently
described by non-local degrees of freedom with a finite number of
terms in the action. Now one could argue that, since one is only interested
in the behavior of the leading Green functions, as far as the appearing
higher order Green functions in the corresponding DSEs are known (e.g.
the 5- \& 6-point functions in Yang-Mills theory) the system for the
primitively divergent Green functions may be solved independent of
these problems. The important point to realize is that in such a case
there is no way to obtain them, neither in the context of functional
techniques like DSEs nor in any other scheme like lattice gauge theory
since even there the number of lattice points corresponds to the highest
Green function that can be realized and that could contribute to the
dynamics of the system in this truncation. Correspondingly, in contrast
to the usual situation where one expects that the precise form of
the unknown higher Green functions becomes irrelevant at sufficiently
high order and thereby a sensible ansatz should be given by general
arguments like symmetry restrictions, this is intrinsically not fulfilled
in case of an explicitly divergent skeleton expansion. Finding a reasonable
ansatz for these higher order Green functions is thereby equivalent
to guessing the correct solution for the primitively divergent Green
functions in the first place. Finally, even if we would simply by
chance guess the precise solution for the required higher Green functions
to solve the system for the lowest correlators like the propagators,
the above line of reasoning shows that these local Green functions
would have nothing to do with the actual physics of the system since
these degrees of freedom should not be suitable to describe the system
in case of a divergent skeleton expansion. From this point of view
we regard the existence of a stable skeleton expansion as a rather
physical requirement for \emph{any} analysis in terms of underlying
local degrees of freedom.
\end{itemize}

\section{Conclusions}

We have presented a power counting formalism for the analysis of the
non-perturbative IR behavior of field theories in the general case
when finite scales are present. This allows the inclusion of kinematic
divergences and dynamical mass generation in the power counting analysis.
In this framework we studied the IR regime of Landau gauge Yang-Mills
theory in more detail and found that the fixed point structure is
more diverse than previously assumed. As a general result that does
not rely on any approximations we find that the DSEs directly exclude
a class of IR fixed points where the IR strength arises directly from
the gluon dynamics. Instead there are two other qualitatively different
solutions for the fixed point behavior.\\
In the \emph{scaling} solution the IR regime is strongly dominated
by the ghost dynamics, as predicted e.g. in \cite{vonSmekal:1997is,Zwanziger:2001kw,Lerche:2002ep,Zwanziger:2003cf,Alkofer:2004it,Fischer:2006vf,Huber:2007kc}.
The structure of the scaling fixed point established in these studies,
however, has to be amended by additional kinematic singularities.
The presence of these singularities is not only consistent with the
uniform scaling rules but has both conceptual and quantitative impact
on the structure of this solution. In a companion article \cite{Alkofer:2008dt}
we present detailed analytic results for the 3-point vertices that
give the complete kinematic dependence and show precisely the same
kinematic divergences found here by pure power counting arguments.
The knowledge about the leading dynamical contributions obtained here
in combination with the analytic results for the 3-point functions
should allow to give an improved value for the IR exponent $\kappa$
in an approximation that treats the 3-point vertices dynamically.
This should include the main dynamical contributions for a precise
prediction of this pertinent parameter.\\
The non-perturbative analysis of kinematic divergences established
here is even more important in the case of QCD. There non-perturbatively
enhanced, strong kinematic divergences of the quark-gluon vertex in
the corresponding scaling solution can provide a description of crucial
aspects of QCD, like the linear rising potential between static color
sources in the quenched theory \cite{Alkofer:2008tt} and its long-range
screening in the dynamical case\cite{Schwenzer:2008vt} as well as
the U(1) anomaly \cite{Alkofer:2008et}. \\
Depending on the renormalization prescription there is another
\emph{decoupling} fixed point were the gluon acquires a mass and decouples
\cite{Boucaud:2008ji,Aguilar:2008xm,Fischer:2008uz}. Current lattice
results for the propagators clearly favor this possibility \cite{Bowman:2007du}.
Yet, we find that, aside from possible inherent problems of these
analyses, cf. \cite{Sternbeck:2005tk,Maas:2008ri,Sternbeck:2008mv},
in the decoupling scenario the gluonic vertices remain bare in the
IR limit and thereby this fixed point shows no infrared enhancement
at all. Correspondingly it does not provide a description of the above
vital aspects of QCD in terms of the Green functions of the fundamental
local degrees of freedom\emph{ alone}. Despite its apparently simple
form it would instead require a more complicated description of the
QCD vacuum in terms of other degrees of freedom.
\begin{acknowledgments}
It is a pleasure to thank Christian Fischer, Felipe Llanes-Estrada,
Axel Maas, Jan Pawlowski and Lorenz von Smekal for enlightening discussions.
K.S. acknowledges support from the Austrian science fund (FWF) under
contract M979-N16, R.A. from the DFG under contract AL 279/5-2 and
M.Q.H. is supported by the Doktoratskolleg ``Hadrons in Vacuum, Nuclei
and Stars'' of the FWF under contract W1203-N08. 
\end{acknowledgments}
\appendix

\section{Graphical derivation of Dyson-Schwinger equations\label{sec:DSE-derivation}}

\begin{figure}
\includegraphics[scale=0.5]{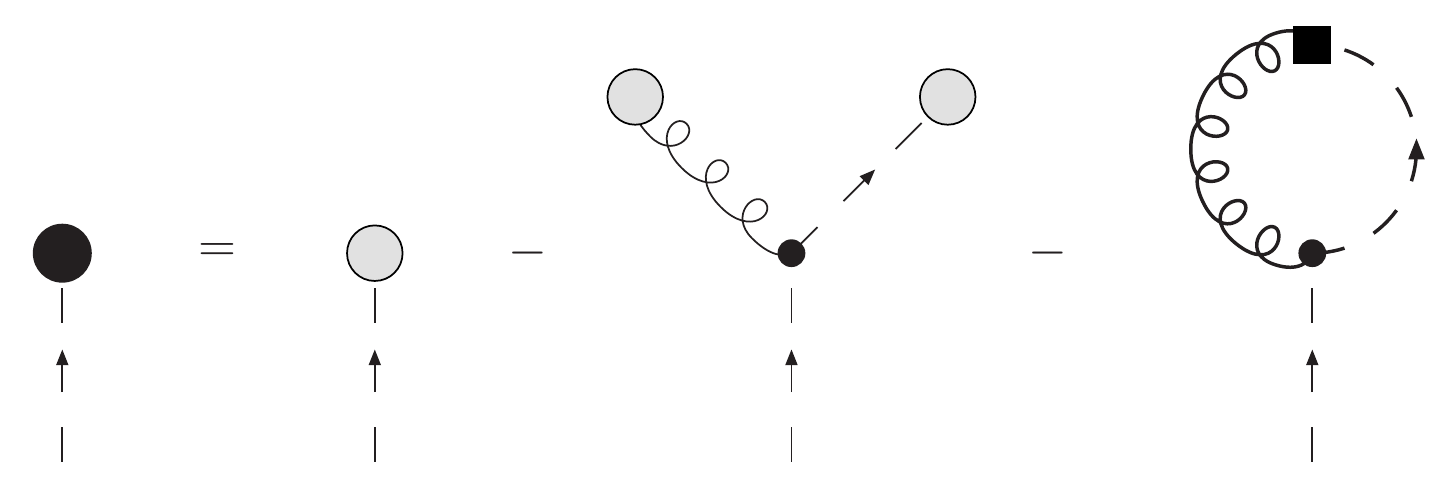}

\caption{\label{fig:generating-ghost-DSE}The generating ghost-DSE. Thin and
thick lines and solid dots represent bare and proper propagators and
vertices. Open dots represent explicit fields and the black square
separating the two different lines denotes the corresponding off-diagonal
component of the super-propagator.}

\end{figure}

\begin{figure}
\includegraphics[scale=0.5]{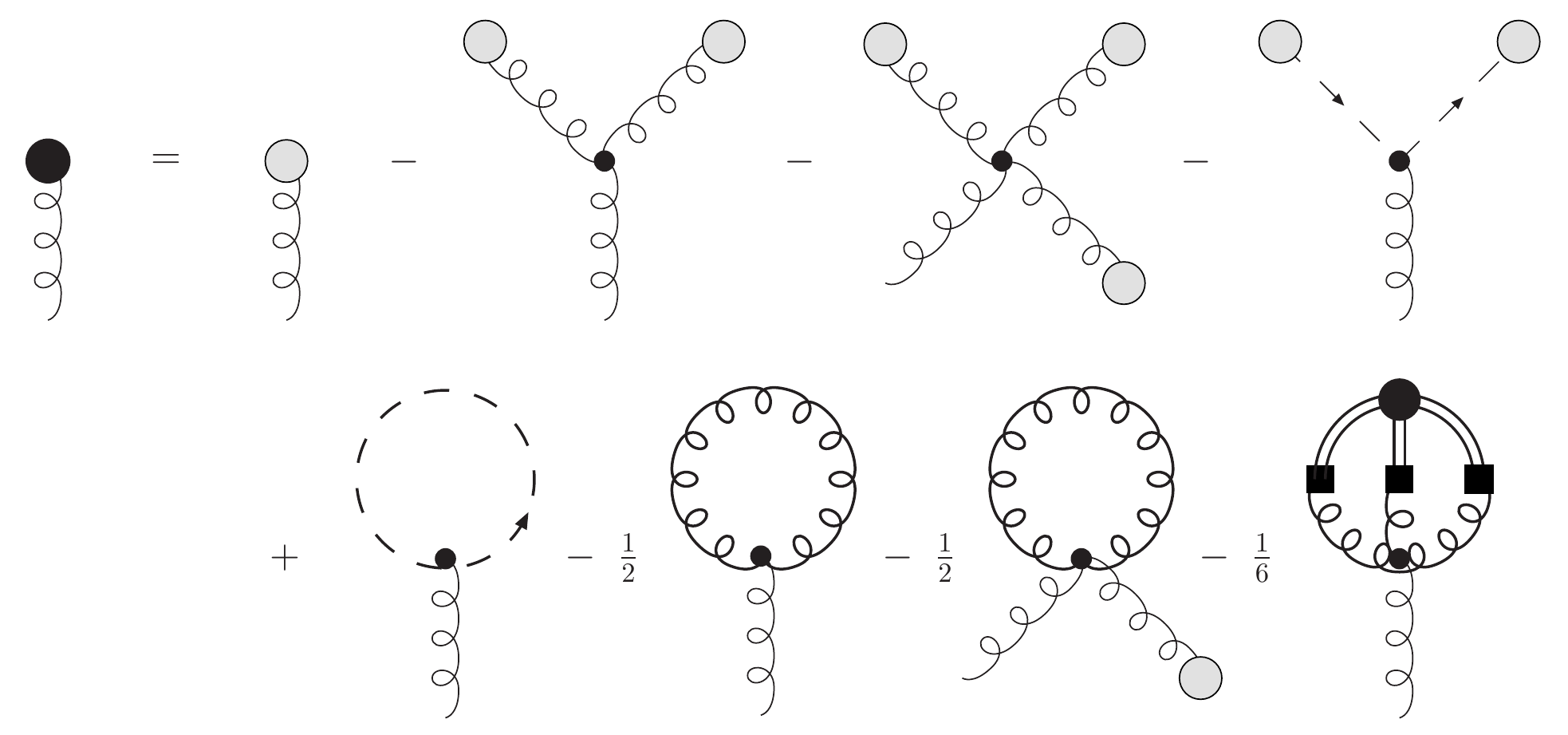}

\caption{\label{fig:generating-gluon-DSE}The generating gluon-DSE. The lhs
in the generating equations represents the field derivative of the
effective action.}

\end{figure}

\begin{figure}[h]
\begin{center}\includegraphics[scale=0.5]{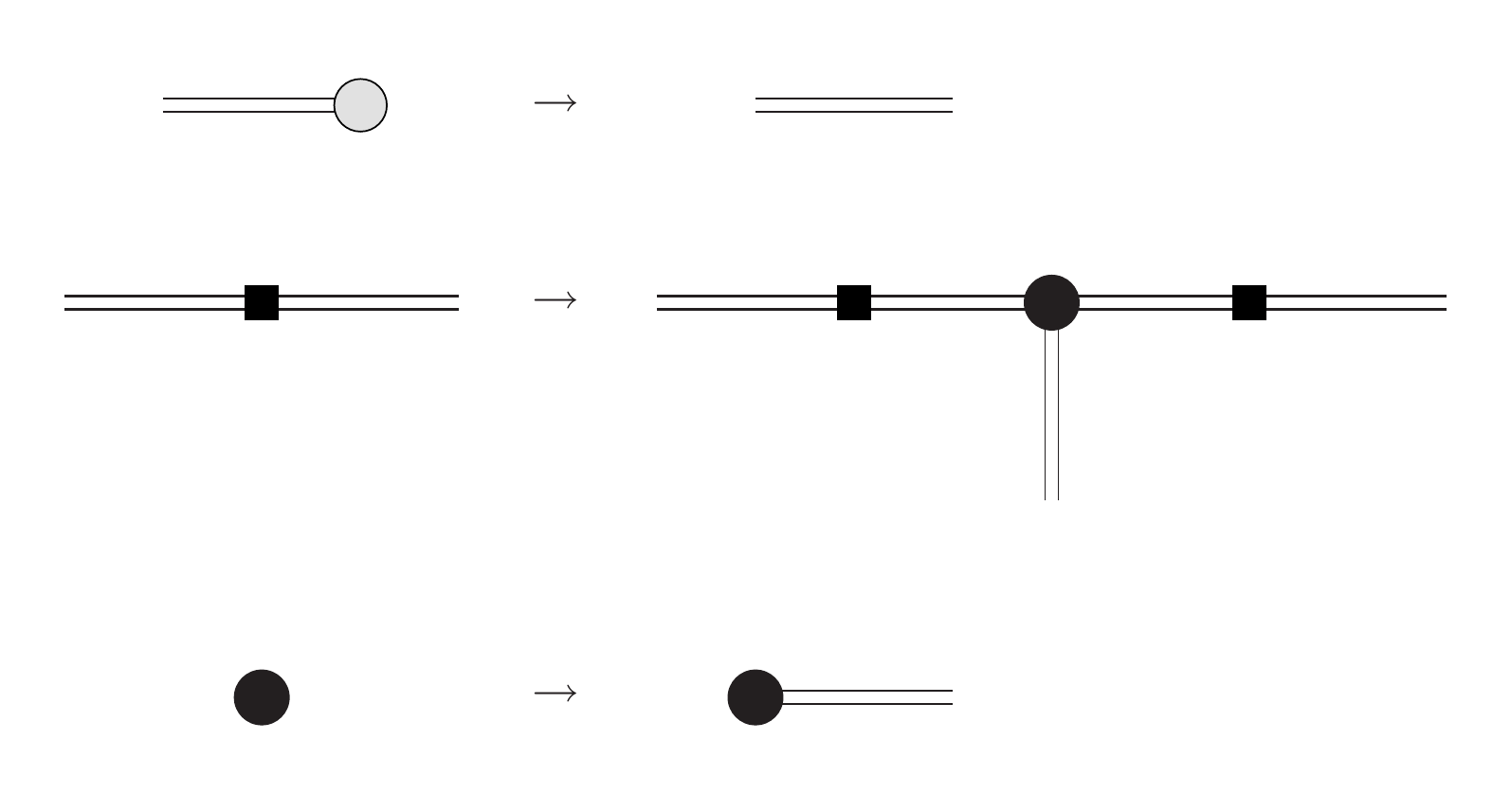}\end{center}

\caption{\label{fig:replacement-rules}The replacement rules for the generation
of DSEs for general correlation functions. One replacement step corresponds
to a functional derivative that generates the DSE for the corresponding
correlation function with one more external leg. In such a step one
of the objects on the lhs is replaced by the corresponding expression
on the rhs. Namely, an explicit field is simply removed. A general
off-shell propagator is extended introducing a new proper 3-point
vertex where each double line can stand for any field in the super-multiplet.
A general proper vertex represented by the thick dot (which can already
have any number of external legs) is simply extended via attaching
another leg.}

\end{figure}

The Dyson-Schwinger equations for the vertices given in the main text
were derived via an algorithmic method presented in \cite{Alkofer:2008nt}.
In this appendix we sketch this method to derive Dyson-Schwinger equations
for general correlation functions from the corresponding equation
for the 1-point function which we will in the following refer to as
the {}``generating DSE''. They involve except for one bare vertex
only proper correlation functions and are given for the ghost and
gluon in figs. \ref{fig:generating-ghost-DSE} and \ref{fig:generating-gluon-DSE},
respectively. As shown in detail in \cite{Alkofer:2008nt} within
a convenient superfield formalism all other DSEs can be computed from
these equations via the replacement rules in fig. \ref{fig:replacement-rules}
where the double lines stand for superfields that include all elementary
fields in the theory. As the final step all expressions have to be
evaluated at their vacuum expectation value which corresponds graphically
to replacing all super-propagators and vertices by the irreducible
ones in all possible ways that involve only physical propagators and
vertices in accordance with the symmetry of the action. This generally
removes many graphs, so that when performing this procedure it is
useful to take into account beforehand to what order the DSEs shall
be computed to neglect any unphysical terms that would already vanish
anyhow during the extension steps.\\
In the special case of non-Abelian gauge theory the super-multiplet
containing gluon and (anti-)ghost fields is given by $\phi=\left(A,\bar{c},c\right)$
and denoted by curly and dashed lines respectively.  Using the general
graphical replacement rules fig. \ref{fig:replacement-rules} yields
in a straightforward way the equations for the leading correlation
functions figs. \ref{fig:ghost-DSE} to \ref{fig:4-gluon-DSE}. The
proper symmetry factors arise simply from different ways of obtaining
the same graph in the above replacements steps. The arising super-propagators
in the 2-loop term of the generating gluon-DSE fig. \ref{fig:generating-gluon-DSE}
only become relevant for the equation for the ghost-gluon vertex and
corresponding higher order correlation functions. For the derivation
of other correlation functions they may be replaced by ordinary gluon
propagators. It is important to note, however, that it is not possible
to derive the equation for the gluonic vertices by additional derivatives
of the propagator equation fig. \ref{fig:gluon-DSE} since one would
miss additional loop contributions from the next to last graph in
fig. \ref{fig:generating-gluon-DSE} that guarantee the symmetry of
the corresponding equations in a perturbative approximation. Finally
we remark, that since there are no fundamental quark-ghost interactions
in QCD the graphical expressions for the generating quark DSE and
the equations for the leading correlation functions are identical
to the corresponding ghost equations displayed in fig. \ref{fig:generating-ghost-DSE}
respectively figs. \ref{fig:ghost-DSE}, \ref{fig:ghost-gluon-DSE-1}
and \ref{fig:ghost-gluon-DSE-2}.

\section{Decomposition of the loop integrals\label{sec:Decomposition}}

\subsection{Massless integrals}

\begin{figure}
\begin{minipage}[t]{0.5\columnwidth}%
\includegraphics[scale=0.75]{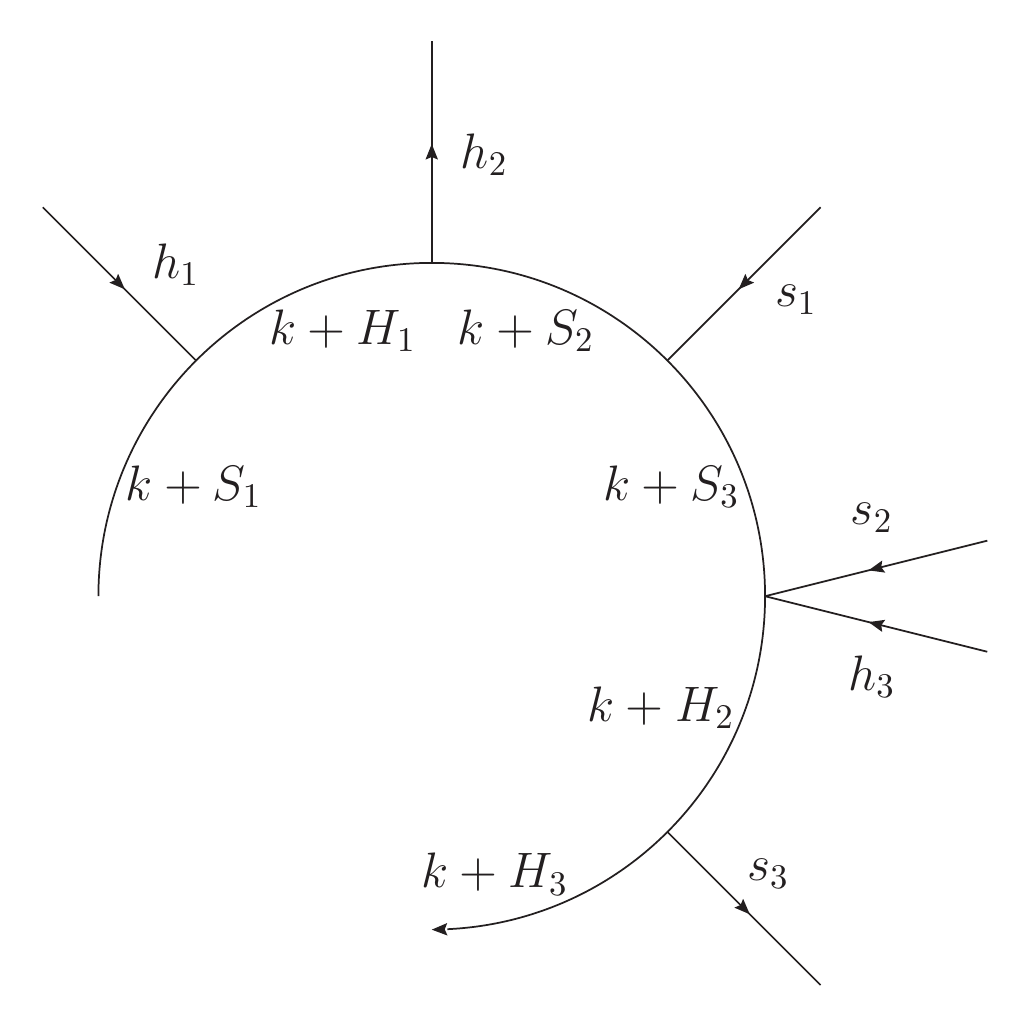}%
\end{minipage}%
\begin{minipage}[t]{0.5\columnwidth}%
\end{minipage}

\caption{Exemplary momentum routing for a general $n$-point function.\label{fig:Momentum-routing}
Here the arrows denote the momentum flow and the particular linear
combinations arising in the diagram are given by $S_{1}=0$, $H_{1}=h_{1}$,
$S_{2}=h_{1}-h_{2}\ll h_{1},h_{2}$, $S_{3}=h_{1}-h_{2}+s_{1}$, $\cdots$.}

\end{figure}

In this appendix we demonstrate explicitly how the loop integrals
in the DSEs can be decomposed into different parts when both soft
and hard external scales are present as the IR limit is taken. A general
1-loop graph is shown in fig. \ref{fig:Momentum-routing}. The external
momenta $q_{n}$ are assumed to be divided into two disjunct subsets
$\{q_{n}\}=\{s_{i}\}\cup\{h_{j}\}$. The $s_{i}$ are soft and the
$h_{j}$ hard, so that in the IR limit $\left|s_{i}\right|\ll\left|h_{j}\right|\:\forall\, i,j$.
Similarly the linear combinations that can be formed out of these
momenta and arise in the integral lines of a general loop graph,

\begin{equation}
Q_{n}=\sum_{i_{n}}\zeta_{i_{n}}q_{i_{n}}\,,\end{equation}
with $\zeta_{i}\in\{-1,0,1\}$, can likewise be divided into corresponding
subsets $\{Q_{n}\}=\{S_{i}\}\cup\{H_{j}\}$ with $\left|S_{i}\right|\ll\left|H_{j}\right|\:\forall\, i,j$.
The integral ${\cal I}$ for a particular term in the general loop
integral ${\cal G}$, eq. (\ref{eq:general-integral}), arising from
the expansion of the product of decomposed vertices eq. (\ref{eq:general-scaling})
reads then

\begin{equation}
{\cal I}=\int\!\!\frac{d^{d}k}{\left(2\pi\right)^{d}}{\cal K}_{\{l_{v}\}}\left(\left\{ \left(k+Q\right)^{2}\right\} ,\left\{ Q^{2}\right\} \right)\prod_{n}\left(k+Q_{n}\right)^{2\left(\delta_{n}-1\right)}\prod_{a}p_{l_{a}}^{2}\left(\left\{ \left(k+Q\right)^{2}\right\} ,\left\{ Q^{2}\right\} \right)^{\delta_{a}^{l_{a}}}\end{equation}
where $n$ runs over the propagators and $a$ over the vertices in
the graph. Here we skip the indices when denoting the corresponding
sets arising as arguments of the functions ${\cal K}$ and $p$ defined
in eqs. (\ref{eq:general-integral}) and (\ref{eq:general-scaling}).
The contributions to the vertex functions with different IR limits
in the decomposition eq. (\ref{eq:general-scaling}) can be divided
into a subset where the power law contains hard external momenta and
another that does not

\begin{align}
{\cal I} & =\int\!\!\frac{d^{d}k}{\left(2\pi\right)^{d}}{\cal K}_{\{l_{v}\}}\left(\left\{ \left(k+S\right)^{2}\right\} ,\left\{ S^{2}\right\} ,\left\{ \left(k+H\right)^{2}\right\} ,\left\{ H^{2}\right\} \right)\prod_{i}\left(k+S_{i}\right)^{2\left(\delta_{i}-1\right)}\prod_{j}\left(k+H_{j}\right)^{2\left(\delta_{j}-1\right)}\nonumber \\
 & \qquad\qquad\qquad\cdot\prod_{a}p_{l_{a}}^{2}\left(\left\{ \left(k+S\right)^{2}\right\} \right)^{\delta_{a}^{l_{a}}}\prod_{b}p_{l_{b}}^{2}\left(\left\{ \left(k+S\right)^{2}\right\} ,\left\{ \left(k+H\right)^{2}\right\} \right)^{\delta_{b}^{l_{b}}}\,.\end{align}
In the IR limit one can choose some intermediate scale $\Theta$,
with $S\ll\Theta\ll H$ where $S\equiv\max\left(\left|S_{i}\right|\right)$
and $H\equiv\min\left(\left|H_{j}\right|\right)$, which can e.g.
be chosen as $\Theta\equiv\sqrt{SH}$. Then the loop integral can
be split into two parts

\begin{align}
{\cal I} & =\left(\int_{0}^{\Theta}\! dk+\int_{\Theta}^{\infty}\! dk\right)k^{d-1}\int\!\!\frac{d\Omega_{d}}{\left(2\pi\right)^{d}}{\cal K}_{\{l_{v}\}}\left(\cdots\right)\prod_{i}\left(k+S_{i}\right)^{2\left(\delta_{i}-1\right)}\prod_{j}\left(k+H_{j}\right)^{2\left(\delta_{j}-1\right)}\prod_{a}p_{l_{a}}^{2}\left(\cdots\right)^{\delta_{a}^{l_{a}}}\prod_{b}p_{l_{b}}^{2}\left(\cdots\right)^{\delta_{b}^{l_{b}}}\nonumber \\
 & \equiv{\cal I}_{<}+{\cal I}_{>}\,.\end{align}
Consider first ${\cal I}_{<}$, where the factors involving the $H_{j}\gg\Theta$
can be expanded in $k$, $S_{i}$

\begin{align}
{\cal I}_{<} & ={\cal K}_{\{l_{v}\}}\left(\left\{ H^{2}\right\} \right)\prod_{j}H_{j}^{2\left(\delta_{j}-1\right)}\prod_{b}p_{l_{b}}^{2}\left(\left\{ H^{2}\right\} \right)^{\delta_{b}^{l_{b}}}\nonumber \\
 & \qquad\cdot\int_{0}^{\Theta}\! dk\, k^{d-1}\int\!\!\frac{d\Omega_{d}}{\left(2\pi\right)^{d}}\prod_{i}\left(k+S_{i}\right)^{2\left(\delta_{i}-1\right)}\prod_{a}p_{l_{a}}^{2}\left(\left\{ \left(k+S\right)^{2}\right\} \right)^{\delta_{a}^{l_{a}}}\left(1+\sum_{j}O\left(\frac{\left|k\right|}{\left|H_{j}\right|}\right)\right)\,.\end{align}
Here the analyticity of the $p_{i}$ as well as the partial analyticity
of ${\cal K}$ has been used. Note that possible non-analyticities
of ${\cal K}_{\{l_{v}\}}$ at hard momenta are no problem as far as
the external momenta $q_{i}$ of the Green function themselves do
not represent a singular configuration at hard external momenta. As
discussed in the main text such a physical pole occurs in Euclidean
space away from the real axis and directly at such a point the solution
of the Dyson-Schwinger equations is not well defined anyhow and we
can therefore simply exclude this special kinematic case here.\\
The integral can now be extended over all scales. The leading part
is

\begin{equation}
{\cal I}_{<}={\cal K}_{\{l_{v}\}}\left(\left\{ H^{2}\right\} \right)\prod_{j}H_{j}^{2\left(\delta_{j}-1\right)}\prod_{b}p_{l_{b}}^{2}\left(\left\{ H^{2}\right\} \right)^{\delta_{b}^{l_{b}}}\int_{0}^{\infty}\! dk\left(\cdots\right)-\Delta{\cal I}_{<}+\tilde{O}\left(\frac{\Theta}{H}\right)\,,\label{eq:Int-low}\end{equation}
where here and in the following the symbol $\tilde{O}$ means that
subleading terms are of the given \emph{relative} order and correspondingly
suppressed compared to the leading terms that are given explicitly.
In the correction term the other factors can be expanded\begin{equation}
\Delta{\cal I}_{<}={\cal K}_{\{l_{v}\}}\left(\left\{ H^{2}\right\} \right)\prod_{j}H_{j}^{2\left(\delta_{j}-1\right)}\prod_{b}p_{l_{b}}^{2}\left(\left\{ H^{2}\right\} \right)^{\delta_{b}^{l_{b}}}\int_{\Theta}^{\infty}\negthickspace dk\int\!\!\frac{d\Omega_{d}}{\left(2\pi\right)^{d}}k^{2\left(\sum_{i}\left(\delta_{i}-1\right)+\frac{d-1}{2}\right)}\prod_{a}p_{l_{a}}^{2}\left(k^{2}\right)^{\delta_{a}^{l_{a}}}\:\left(1\!+\! O\left(\frac{\left|S_{i}\right|}{\left|k\right|}\right)\right).\end{equation}
 Analogously one can expand the upper part of the integral

\begin{align}
{\cal I}_{>} & =\int_{\Theta}^{\infty}\! dk\int\!\!\frac{d\Omega_{d}}{\left(2\pi\right)^{d}}{\cal K}_{\{l_{v}\}}\left(\left\{ \left(k+H\right)^{2}\right\} ,\left\{ H^{2}\right\} \right)k^{2\left(\sum_{i}\left(\delta_{i}-1\right)+\frac{d-1}{2}\right)}\prod_{j}\left(k+H_{j}\right)^{2\left(\delta_{j}-1\right)}\nonumber \\
 & \qquad\cdot\prod_{a}p_{l_{a}}^{2}\left(k^{2}\right)^{\delta_{a}^{l_{a}}}\prod_{b}p_{l_{b}}^{2}\left(k^{2},\left\{ \left(k+H\right)^{2}\right\} \right)^{\delta_{b}^{l_{b}}}\left(1+\sum_{i}O\left(\frac{\left|S_{i}\right|}{\left|k\right|}\right)\right)=\int_{0}^{\infty}\! dk\left(\cdots\right)-\Delta{\cal I}_{>}+\tilde{O}\left(\frac{S}{\Theta}\right)\end{align}
and subsequently the correction term so that their sum gives\begin{align}
\Delta{\cal I} & =\Delta{\cal I}_{<}+\Delta{\cal I}_{>}\\
 & ={\cal K}_{\{l_{v}\}}\left(\left\{ H^{2}\right\} \right)\prod_{j}H_{j}^{2\left(\delta_{j}-1\right)}\prod_{b}p_{b}^{2}\left(\left\{ H^{2}\right\} \right)^{\delta_{b}}\int_{0}^{\infty}\! dk\int\!\!\frac{d\Omega_{d}}{\left(2\pi\right)^{d}}k^{2\left(\sum_{i}\left(\delta_{i}-1\right)+\frac{d-1}{2}\right)}\prod_{a}p_{l_{a}}^{2}\left(k^{2}\right)^{\delta_{a}^{l_{a}}}+\tilde{O}\left(\frac{S}{\Theta},\frac{\Theta}{H}\right)\,.\nonumber \end{align}
This integral involves no scale and thereby simply represents a, possibly
infinite, constant multiplying its prefactor. This prefactor is in
particular completely independent of soft scales and $\Delta{\cal I}$
does thereby not affect the IR power laws of the corresponding Green
functions. In the case of primitively divergent Green functions considered
in this work such a constant contribution is already present in the
DSEs in the form of the tree level term. Moreover, for primitively
divergent Green functions the integrals $\Delta{\cal I}$ can be explicitly
UV divergent. Such a divergence cancels by construction precisely
the additional UV divergence introduced in the additional scale dependent
integral via eq. (\ref{eq:Int-low}). Correspondingly, this constant
contribution could be formally absorbed into the multiplicative renormalization
factor of the tree level term, which thereby has the form to cancel
all divergences in both scale dependent integrals. We note that e.g.
in dimensional renormalization the above scale independent integral
vanishes by definition. Thereby, after proper renormalization the
initial integral is split into two finite contributions that are in
the IR limit $S\ll H$ independent of the scale $\Theta$

\begin{align}
 & {\cal I}={\cal K}_{\{l_{v}\}}\!\left(\left\{ H^{2}\right\} \right)\prod_{j}H_{j}^{2\left(\delta_{j}-1\right)}\prod_{b}p_{l_{b}}^{2}\,\left(\left\{ H^{2}\right\} \right)^{\delta_{b}^{l_{b}}}\int\!\!\frac{d^{d}k}{\left(2\pi\right)^{d}}\prod_{i}\left(k\!+\! S_{i}\right)^{2\left(\delta_{i}\!-\!1\right)}\prod_{a}p_{l_{a}}^{2}\!\left(\left\{ \left(k\!+\! S\right)^{2}\right\} \right)^{\delta_{a}^{l_{a}}}\label{eq:decomposition}\\
 & +\!\int\!\!\frac{d^{d}k}{\left(2\pi\right)^{d}}{\cal K}_{\{l_{v}\}}\!\left(\left\{ \left(k\!+\! H\right)^{2}\right\} ,\left\{ H^{2}\right\} \right)\prod_{j}\left(k\!+\! H_{j}\right)^{2\left(\delta_{j}\!-\!1\right)}\prod_{b}p_{l_{b}}^{2}\!\left(k^{2},\left\{ \left(k\!+\! H\right)^{2}\right\} \right)^{\delta_{b}^{l_{b}}}\! k^{2\sum_{i}\left(\delta_{i}\!-\!1\right)}\prod_{a}p_{l_{a}}^{2}\!\left(k^{2}\right)^{\delta_{a}^{l_{a}}}\negmedspace+\!\tilde{O}\left(\frac{S}{H}\right).\nonumber \end{align}
The first integral depends now only on soft external momenta, whereas
the second one depends only on hard momenta. These integrals are generally
dominated by the poles of the integrand and have to scale with the
respective external scales. Correspondingly once the contributions
from scales of different order of magnitude are separated their scale
dependence can be analyzed by mere power counting.\\
Yet, the hard integral can still be subject to soft singularities
when differences of the hard momenta become soft. Then the formally
absent soft momenta re-enter via the condition of momentum conservation.
To see this, consider a 3-point integral with one soft and two hard
external momenta. For better readability we skip in the following
the vertex dressing functions and the kernel ${\cal K}$ but note
that the derivation proceeds analogously with these complications.
The decomposition eq. (\ref{eq:decomposition}) reads then

\begin{align}
{\cal I}^{\left(3\right)} & =h^{2\left(\alpha+\beta-2\right)}\!\int\!\!\frac{d^{d}k}{\left(2\pi\right)^{d}}\left(k-s\right)^{2\left(\gamma-1\right)}+\int\!\!\frac{d^{d}k}{\left(2\pi\right)^{d}}\left(k+h-s\right)^{2\left(\alpha-1\right)}\left(k+h\right)^{2\left(\beta-1\right)}k^{2\left(\gamma-1\right)}+\tilde{O}\left(\frac{\left|s\right|}{\left|h\right|}\right)\,.\end{align}
The momentum combination $k+h-s$ in the hard part of the loop integral
is again sensitive to the soft momentum $s$ in a small region around
the loop momentum $k=-h$. To conveniently study this region it is
useful to shift the integration variable via the substitution $k^{\prime}=k+h$
in order to divide the integral at some scale $\Theta$

\begin{equation}
{\cal I}_{>}^{\left(3\right)}=\!\int\!\!\frac{d^{d}k^{\prime}}{\left(2\pi\right)^{d}}\left(k^{\prime}\!-\! s\right)^{2\left(\alpha-1\right)}k^{\prime2\left(\beta-1\right)}\left(k^{\prime}\!-\! h\right)^{2\left(\gamma-1\right)}=\left(\int_{0}^{\Theta}\negthickspace dk^{\prime}\!+\!\int_{\Theta}^{\infty}\negthickspace dk^{\prime}\right)\!\int\!\!\frac{d\Omega_{d}}{\left(2\pi\right)^{d}}\left(k^{\prime}\!-\! s\right)^{2\left(\alpha-1\right)}k^{\prime2\left(\beta+\frac{d-3}{2}\right)}\left(k^{\prime}\!-\! h\right)^{2\left(\gamma-1\right)}.\end{equation}
In complete analogy to the general case above (using $s_{1}=s$, $s_{2}=0$,
$h_{1}=-h$ and $\delta_{i=1}=\alpha$, $\delta_{i=2}=\beta$, $\delta_{j=1}=\gamma$)
this integral can then be decomposed once more so that the full 3-point
integral is effectively divided into three distinct integrals extending
over all scales

\begin{equation}
{\cal I}^{\left(3\right)}=h^{2\left(\alpha+\beta-2\right)}\negthickspace\int\!\!\frac{d^{d}k}{\left(2\pi\right)^{d}}\left(k\!-\! s\right)^{2\left(\gamma-1\right)}+h^{2\left(\gamma-1\right)}\negthickspace\int\!\!\frac{d^{d}k}{\left(2\pi\right)^{d}}\left(k\!-\! s\right)^{2\left(\alpha-1\right)}k^{2\left(\beta-1\right)}+\int\!\!\frac{d^{d}k}{\left(2\pi\right)^{d}}k^{2\left(\alpha+\beta-2\right)}\left(k\!+\! h\right)^{2\left(\gamma-1\right)}+\tilde{O}\left(\frac{\left|s\right|}{\left|h\right|}\right).\end{equation}
The first two integrals are, up to $\tilde{O}\left(\left|s\right|/\left|h\right|\right)$
corrections, completely IR dominated whereas the third one is dominated
by hard scales. This division is graphically displayed in fig. \ref{fig:soft-loops}
in the main text by distinct loop graphs with color-coded soft and
hard propagators. The case of 2-loop contributions in the DSEs can
be handled accordingly, with the complication that both loop integrals
have to be split and the number of distinct integrals increases. In
general the number of possible divisions depends on the $n$-point
function, the number of loops and the numbers of soft and hard external
momenta. After a full decomposition of the integrals there are -
aside from purely hard integrals that yield a scale independent constant
- only integrals left that depend on a given set of soft external
momenta $\{s_{i}\}$. These momenta can be parameterized in the form
$s_{i}=\rho_{i}\, s$ with a single scaling variable $s$ and fixed,
bounded momentum ratios $\rho_{i}\equiv s_{i}/s$, $\left|\rho_{i}\right|\leq1$
that could likewise be replaced by angular variables. In such a case
the qualitative IR scaling of the corresponding integral is independent
of the $\rho_{i}$ and can be determined by a pure power counting
analysis. The prefactor of the IR scaling law however involves the
ratios $\rho_{i}$ and requires an explicit solution of the integral
\cite{Alkofer:2008dt}. For the cases we study in this work, the
integrals that need to be decomposed depend even on a single soft
scale and therefore trivially scale as some power of it.

\subsection{Massive integrals}

In this subsection we extend the previous decomposition of loop integrals
to the case when mass scales are involved. As already discussed in
the main text, in the case that all propagators in an integral are
massive one can in the IR limit $p\ll m$ simply expand the whole
integral, e.g.

\begin{equation}
\int d^{d}k\frac{1}{\left(k+p\right)^{2}+m^{2}}\frac{1}{k^{2}+m^{2}}\approx\int d^{d}k\frac{1}{\left(k^{2}+m^{2}\right)^{2}}+\tilde{O}\left(\frac{\left|p\right|}{m}\right)\end{equation}
to convince oneself that the integral does not depend on the soft
external scale to leading order. Even though there is no pole here
the integral is dominated by modes of the order of the mass: the contribution
from low modes is suppressed by phase space and the renormalized contribution
from large modes is suppressed by the denominators.\\
When there are both massive and massless propagators present the
integral receives contributions from both soft and hard scales. To
see this consider the massive $2$-point integral of the form

\begin{equation}
{\cal I}_{m}\equiv\int d^{d}k\frac{1}{\left(k+p\right)^{2\alpha}}\frac{1}{k^{2}+m^{2}}\:.\end{equation}
In the limit $p\ll m$ one can again introduce an arbitrary intermediate
scale $\Theta$ and divide the integrals

\begin{align}
{\cal I}_{m} & =\left(\int_{0}^{\Theta}dk\, k^{d-1}\int d\Omega+\int_{\Theta}^{\infty}dk\, k^{d-1}\int d\Omega\right)\frac{1}{\left(k+p\right)^{2\alpha}}\frac{1}{k^{2}+m^{2}}\nonumber \\
 & =\frac{1}{m^{2}}\int_{0}^{\Theta}dk\, k^{d-1}\int d\Omega\frac{1}{\left(k+p\right)^{2\alpha}}\left(1+O\left(\frac{\left|k\right|}{m}\right)\right)+\int_{\Theta}^{\infty}dk\, k^{d-1}\int d\Omega\frac{1}{k^{2\alpha}}\frac{1}{k^{2}+m^{2}}\left(1+O\left(\frac{\left|p\right|}{\left|k\right|}\right)\right)\nonumber \\
 & =\frac{1}{m^{2}}\int_{0}^{\infty}d^{d}k\frac{1}{\left(k+p\right)^{2\alpha}}+\int_{0}^{\infty}d^{d}k\frac{1}{k^{2\alpha}}\frac{1}{k^{2}+m^{2}}-\frac{1}{m^{2}}\int_{0}^{\infty}d^{d}k\frac{1}{k^{2\alpha}}+\tilde{O}\left(\frac{\left|p\right|}{m}\right)\end{align}
where the scale independent integral again cancels momentum independent
divergences introduced in the decomposition. The first integral is
dominated by soft scales and the second one by hard scales of the
order of the mass and correspondingly the scale separation in the
IR limit allows again the use of a power counting analysis. An analogous
decomposition is possible for massive vertex integrals as well as
for integrals that involve both masses and hard external momentum
scales. In all these cases the presented decomposition results in
a complete separation of the various external scales in different
loop integrals and allows thereby the application of power counting
methods. Finally we note that the presented method for decomposing
massive integrals is likewise suitable for multi-loop corrections,
like the 2-loop graphs arising in certain DSEs, via successive decompositions.

\end{document}